\documentclass[aps,prl,reprint,superscriptaddress,twocolumn,longbibliography]{revtex4-1}
\usepackage{amsfonts,amssymb,amscd,amsthm}
\usepackage{graphicx}
\usepackage{mathrsfs}
\usepackage{soul,color,xcolor}
\usepackage[intlimits]{amsmath}
\usepackage[colorlinks, citecolor=red]{hyperref}
\usepackage{etoolbox}
\usepackage{upgreek}

\begin{document}
\title{Realization of a functioning dual-type trapped-ion quantum network node}

\author{Y.-Y. Huang}
\thanks{These authors contribute equally to this work}%
\affiliation{Center for Quantum Information, Institute for Interdisciplinary Information Sciences, Tsinghua University, Beijing 100084, PR China}

\author{L. Feng}
\thanks{These authors contribute equally to this work}%
\affiliation{Center for Quantum Information, Institute for Interdisciplinary Information Sciences, Tsinghua University, Beijing 100084, PR China}

\author{Y.-K. Wu}
\thanks{These authors contribute equally to this work}%
\affiliation{Center for Quantum Information, Institute for Interdisciplinary Information Sciences, Tsinghua University, Beijing 100084, PR China}
\affiliation{Hefei National Laboratory, Hefei 230088, PR China}

\author{Y.-L. Xu}
\affiliation{Center for Quantum Information, Institute for Interdisciplinary Information Sciences, Tsinghua University, Beijing 100084, PR China}

\author{L. Zhang}
\affiliation{Center for Quantum Information, Institute for Interdisciplinary Information Sciences, Tsinghua University, Beijing 100084, PR China}

\author{Z.-B. Cui}
\affiliation{Center for Quantum Information, Institute for Interdisciplinary Information Sciences, Tsinghua University, Beijing 100084, PR China}

\author{C.-X. Huang}
\affiliation{Center for Quantum Information, Institute for Interdisciplinary Information Sciences, Tsinghua University, Beijing 100084, PR China}

\author{C. Zhang}
\affiliation{HYQ Co., Ltd., Beijing 100176, P. R. China}

\author{S.-A. Guo}
\affiliation{Center for Quantum Information, Institute for Interdisciplinary Information Sciences, Tsinghua University, Beijing 100084, PR China}

\author{Q.-X. Mei}
\affiliation{HYQ Co., Ltd., Beijing 100176, P. R. China}

\author{B.-X. Qi}
\affiliation{Center for Quantum Information, Institute for Interdisciplinary Information Sciences, Tsinghua University, Beijing 100084, PR China}

\author{Y. Xu}
\affiliation{Center for Quantum Information, Institute for Interdisciplinary Information Sciences, Tsinghua University, Beijing 100084, PR China}
\affiliation{Hefei National Laboratory, Hefei 230088, PR China}

\author{Y.-F. Pu}
\affiliation{Center for Quantum Information, Institute for Interdisciplinary Information Sciences, Tsinghua University, Beijing 100084, PR China}
\affiliation{Hefei National Laboratory, Hefei 230088, PR China}

\author{Z.-C. Zhou}
\affiliation{Center for Quantum Information, Institute for Interdisciplinary Information Sciences, Tsinghua University, Beijing 100084, PR China}
\affiliation{Hefei National Laboratory, Hefei 230088, PR China}

\author{L.-M. Duan}
\email{lmduan@tsinghua.edu.cn}
\affiliation{Center for Quantum Information, Institute for Interdisciplinary Information Sciences, Tsinghua University, Beijing 100084, PR China}
\affiliation{Hefei National Laboratory, Hefei 230088, PR China}

\begin{abstract}
Trapped ions constitute a promising platform for implementation of a quantum network. Recently, a dual-type qubit scheme has been realized in a quantum network node where the communication qubits and the memory qubits are encoded in different energy levels of the same ion species, such that the generation of ion-photon entanglement on the communication qubits has negligible crosstalk error on the preloaded quantum information in the memory qubits. However, to achieve the versatile applications of a quantum network, a crucial component of the dual-type node, namely the entangling gate between the communication and the memory qubits, is still missing. Here we report a dual-type quantum network node equipped with ion-photon entanglement generation, crosstalk-free quantum memory and entangling gates between the dual-type qubits simultaneously. We demonstrate its practical applications including the quantum state teleportation and the preparation of multipartite entangled state. Our work achieves the necessary components of a dual-type quantum network node and paves the way toward its applications in a large-scale quantum internet.
\end{abstract}

\maketitle

\section{Introduction}
A quantum network \cite{Kimble2008,doi:10.1126/science.aam9288} plays a vital role in quantum information science and has wide applications in quantum communication \cite{Gisin2007,RevModPhys.81.1301,RevModPhys.92.025002}, precision measurement \cite{doi:10.1126/science.1104149,Komar2014,PhysRevLett.109.070503} and distributed quantum computing \cite{PhysRevA.59.4249,PhysRevA.76.062323,broadbent2009blind}. To enable efficient transmission and reliable storage of the quantum information, a quantum network typically consists of flying photons for communication, and matter qubits encoded in physical systems like trapped ions \cite{10.5555/2011617.2011618,RevModPhys.82.1209}, atomic ensembles \cite{Duan2001DLCZ,RevModPhys.82.1041,RevModPhys.83.33}, cavity QED systems \cite{RevModPhys.87.1379,RevModPhys.94.041003}, NV centers \cite{Childress_Hanson_2013,Awschalom2018} or superconducting circuits \cite{Mirhosseini2020,Zhong2021} for long-time quantum storage. Among these physical systems, trapped ions have the advantages of a long coherence time \cite{wang2021single} and deterministic and high-fidelity gate operations \cite{PhysRevLett.113.220501,PhysRevLett.117.060504,PhysRevLett.117.060505,PhysRevLett.127.130505} for the matter qubits, as well as their high-fidelity entanglement with photons that can transmit in commercial optical fibers \cite{Blinov2004,Hucul2015,PhysRevLett.124.110501,Bock2018,PhysRevLett.130.050803}.

To support versatile functions of a quantum internet, a minimal quantum network node requires two types of matter qubits: one as communication qubits for high-rate generations of matter-light entanglement, and the other as memory qubits which can store the quantum information and can be entangled with the communication qubits by high-fidelity gate operations \cite{doi:10.1126/science.aam9288}. Then with further entanglement swapping between photonic qubits from different quantum network nodes, the matter qubits in these nodes can be entangled altogether for later entanglement purifications or quantum error corrections, and finally for the various applications in quantum information processing \cite{Kimble2008,doi:10.1126/science.aam9288}. However, for trapped ions, the generation of ion-photon entangled pairs naturally accompanies the spontaneous emission of the communication ions and hence the scattered photons in random directions. These photons, being able to resonantly drive transitions of other memory ions, will severely degrade their storage lifetime, erasing any quantum information when waiting for the communication qubits to establish entanglement \cite{Feng2024}. Therefore, in the early demonstrations of a trapped-ion quantum network node, usually the memory qubit has to be reset after the successful entanglement generation on the communication qubit \cite{Hucul2015}, limiting the possible applications of the quantum network node. One possible scheme to solve this problem is to use two different ion species for the communication qubits and the memory qubits, such that the frequency of the emitted photons from the former is no longer resonant to the latter \cite{PhysRevLett.118.250502,PhysRevLett.130.090803}. Following this route, recently distributed quantum algorithms have been demonstrated between two network nodes, each containing a memory qubit and a communication qubit without crosstalk errors \cite{Main2025}.

However, when increasing the qubit number in each node, the dual-species scheme faces the difficulty of a lowered sympathetic cooling efficiency due to the mass mismatch between the ions \cite{PhysRevA.103.012610}. While this can be overcome by the quantum charge-coupled device (QCCD) architecture which breaks a large ion crystal into small pieces with the capability to shuttle individual ions on demand \cite{Kielpinski2002}, existing experiments \cite{you2024temporally,cui2025metropolitan} have not yet shown the entanglement between the communication and the memory qubits. On the other hand, recently a dual-type qubit scheme has been proposed \cite{yang2022realizing,10.1063/5.0069544} and demonstrated \cite{yang2022realizing,Feng2024,PhysRevLett.134.070801} where the two types of the matter qubits can be encoded into different hyperfine levels of the same ion species without crosstalk errors. This scheme is then compatible with the recently developed two-dimensional ion crystal architecture for which hundreds of ionic qubits \cite{guo2024siteresolved} and individually addressed entangling gates \cite{Hou2024} have been achieved. Enabled by the dual-type qubit scheme, a crosstalk-free quantum network node has been realized \cite{Feng2024,cui2025metropolitan} with the memory lifetime exceeding the time required to generate the ion-photon entanglement. However, a fully-functioning dual-type quantum network node, with entangling gates between the communication and the memory qubits to enable various applications, has not yet been demonstrated.

\begin{figure*}[!tp]
   \includegraphics[width=5.1in]{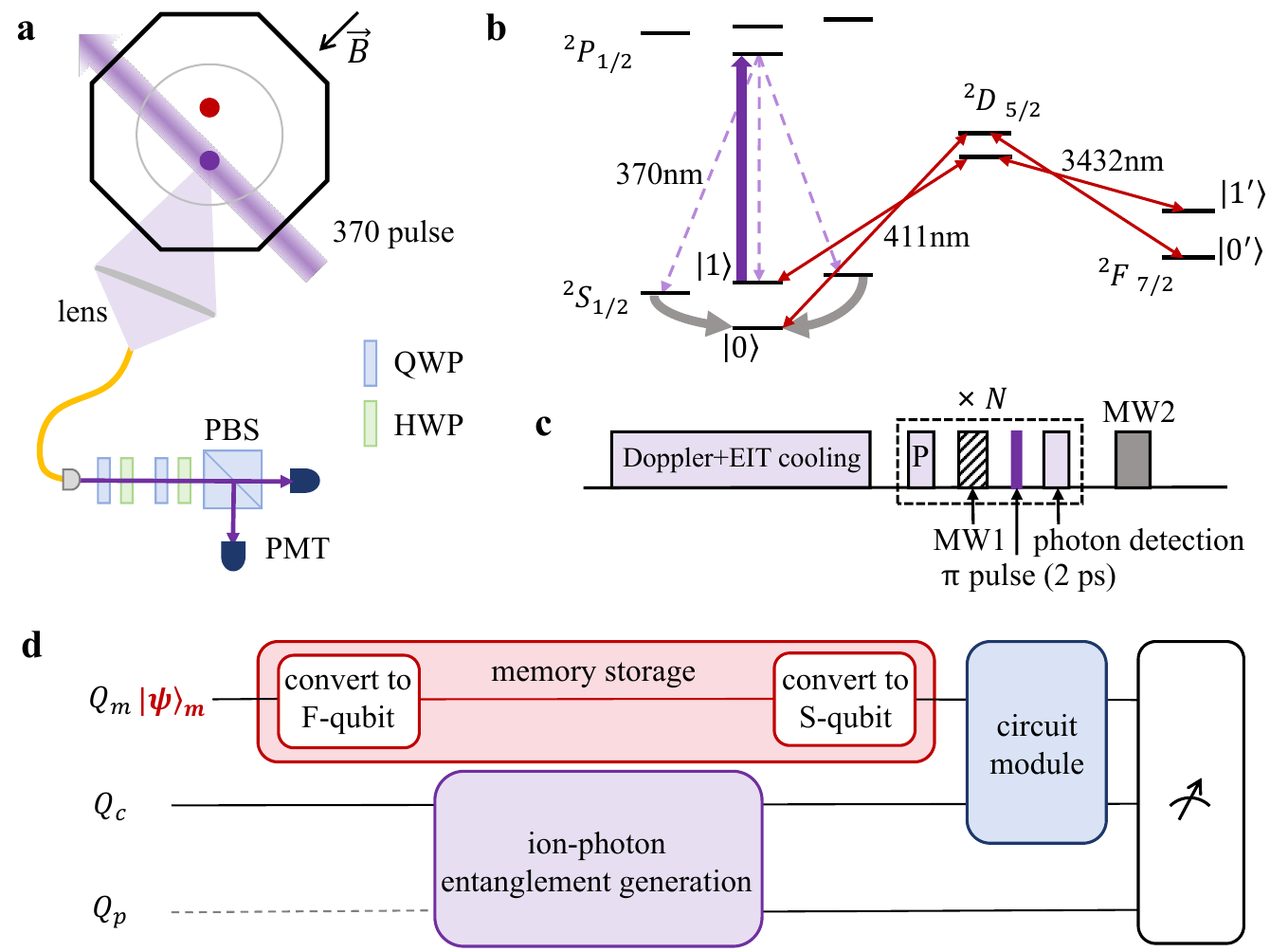}
   \caption{\textbf{Experimental scheme for a two-ion quantum network node under dual-type encoding.}
   \textbf{a}, Two $^{171}$Yb$^+$ ions are trapped with a separation of $10\,\upmu$m. The red ion represents the $F$-type memory qubit, and the purple ion represents the $S$-type communication qubit. When excited by a focused $370\,$nm laser beam, the communication ion can generate ion-photon entanglement. The radiated photon is then collected by a lens (perpendicular to the paper plane, here depicted on the side of the trap for clarity) and is detected by a polarization beam splitter (PBS) and two photomultiplier tubes (PMT) in a basis defined by the quarter-waveplates (QWP, blue) and half-waveplates (HWP, green).
   \textbf{b}, Relevant energy levels of the $^{171}$Yb$^+$ ion. The $S$-type qubit is defined between a pair of clock states in the $^2S_{1/2}$ manifolds, while the $F$-type qubit between another pair of clock states in the $^2F_{7/2}$ manifolds. The qubit type can be coherently converted via the intermediate $^2D_{5/2}$ levels using a focused bichromatic $411\,$nm laser beam and a global bichromatic $3432\,$nm laser beam.
   \textbf{c}, Experimental sequence for ion-photon entanglement generation. After the Doopler cooling and EIT cooling, we use an optical pumping (P) pulse and a microwave $\pi$-pulse (MW1) to initialize the ion in $|1\rangle$. Then a $\pi$-polarized $370\,$nm laser $\pi$-pulse [broad purple arrow in (b)] will excite the ion to $|{}^2 P_{1/2},F=0, m_F=0\rangle$. If a spontaneous-emission photon is detected in a given time window, a two-tone microwave pulse [MW2, gray arrows in (b)] will be applied to prepare the ion-photon EPR state.
   \textbf{d}, Schematic circuit for the experimental protocol. The memory qubit $Q_m$, initialized in $|\psi\rangle_m$, is converted to the $F$-type for storage. After an ion-photon entanglement is generated between the communication qubit $Q_c$ and the photonic qubit $Q_p$, the memory qubit is converted back to the $S$-type for further entanglement operations or Bell measurement with the communication qubit.
   \label{fig:scheme}}
\end{figure*}

Here we report such a fully-functioning dual-type quantum network node. We first calibrate elementary functions of the node, including the generation of ion-photon entanglement, the long-time storage of the memory qubit, the entangling gate between the memory and the communication ion qubits, and the qubit type conversion for the dual-type qubits. Then we utilize the crosstalk-free quantum network node for practical applications from quantum state teleportation to the preparation of a GHZ entangled state between the two matter qubits and one photonic qubit. Our work demonstrates the compatibility of all the necessary elements of a dual-type quantum network node and paves the way toward a large-scale quantum internet.

\section{Results}
\subsection{Experimental scheme}
Our experimental setup is schematically illustrated in Fig.~\ref{fig:scheme}\textbf{a}. Two ${}^{171}\mathrm{Yb}^+$ ions are held in a blade trap at a distance of $10\,\upmu$m. The communication ion can be excited by a focused $370\,$nm laser pulse with a beam waist radius of $4\,\upmu$m to generate the ion-photon entanglement through the spontaneous emission. We use a $0.23\,$NA objective lens in a direction perpendicular to the magnetic field ($B\approx5.46\,$G) to collect the radiated single photon, which is then coupled to a single mode fiber for further detection by a polarization beam splitter and two photomultiplier tubes (PMTs). We use one set of a quarter-waveplate and a half-waveplate to compensate the fiber-induced polarization drift, and then another set to choose the measurement basis of the photon polarization state.

The experimental sequence for the ion-photon entanglement generation is sketched in Fig.~\ref{fig:scheme}\textbf{c}. After $40\,\upmu$s Doopler cooling and $200\,\upmu$s EIT cooling, we use a $5\,\upmu$s optical pumping pulse and a $10\,\upmu$s microwave $\pi$-pulse to initialize the ion in $|1\rangle\equiv |{}^2 S_{1/2},F=1, m_F=0\rangle$. Then we apply a $2\,$ps $\pi$-polarized $370\,$nm laser $\pi$-pulse (broad purple arrow in Fig.~\ref{fig:scheme}\textbf{b}) to excite the ion to $|{}^2 P_{1/2},F=0, m_F=0\rangle$.
Upon the detection of the spontaneous emission photon in a $60\,$ns window, we apply a two-tone microwave pulse as shown by the gray arrows in Fig.~\ref{fig:scheme}\textbf{b} to merge the population on $|{}^2 S_{1/2},F=1, m_F=\pm 1\rangle$ levels into $|0\rangle\equiv |{}^2 S_{1/2},F=0, m_F=0\rangle$, and obtain a desired EPR state between the $S$-type communication qubit and the photonic qubit \cite{Feng2024}
\begin{equation}
|\Phi \rangle_{cp} = \frac{1}{\sqrt{2}}
\left(|0\rangle_{c}|H\rangle_{p} + |1\rangle_{c}|V\rangle_{p}\right), \label{eq:entangled}
\end{equation}
where $|H\rangle$ and $|V\rangle$ represent the polarization of the photonic qubit. If the single photon is not detected in the given time window, we repeat the cycles in the dashed box in Fig.~\ref{fig:scheme}\textbf{c} for up to $N=10$ attempts before we go back to the Doppler and EIT cooling stage again. This choice is to balance the duty cycle of the entanglement generation and the motional heating effect. More details can be found in Supplementary Information.

To avoid the crosstalk from the randomly scattered photons during the ion-photon entanglement generation, we encode the memory qubit $|\psi\rangle_m$ in the $F$-type spanned by $|0^\prime\rangle\equiv |{}^2 F_{7/2},F=3, m_F=0\rangle$ and $|1^\prime\rangle\equiv |{}^2 F_{7/2},F=4, m_F=0\rangle$ levels. The coherent conversion between the $S$-qubit and the $F$-qubit can be achieved by sequentially applying a focused $411\,$nm laser with a beam waist radius of $3\,\upmu$m and a global $3432\,$nm laser via the intermediate $D_{5/2}$ states \cite{yang2022realizing}, as shown by the red arrows in Fig.~\ref{fig:scheme}\textbf{b}.

To further entangle the memory qubit and the communication qubit, we first convert them to the $S$-type and then apply a global $411\,$nm laser beam with a beam waist radius of $35\,\upmu$m to engineer a $\sigma_z$-type spin-dependent force (SDF) \cite{PhysRevA.107.022617,Roos_2008}. The bichromatic $411\,$nm laser has its two tones detuned symmetrically from the $|^2 S_{1/2},F=0, m_F=0\rangle \leftrightarrow |^2 D_{5/2},F=2, m_F=0\rangle$ transition with a detuning $\delta = \pm(\omega_c + \omega_r)/2$, where $\omega_c$ and $\omega_r$ denote the center-of-mass (COM) and the rolling mode frequencies in the radial direction, respectively. To suppress off-resonant excitations, we set a $6\,\upmu$s ramping time in the beginning and the end of the pulse with a $\mathrm{sin}^2$ pulse shape \cite{PhysRevA.107.022617}. As shown in the inset of Fig.~\ref{fig:fidelity}\textbf{c}, we further insert a microwave $\pi$ pulse in the middle of two segments of time evolutions under the SDF Hamiltonian, such that the two qubit states $|00\rangle$ and $|11\rangle$ accumulate the same geometric phase. Overall, this pulse sequence generates an entangling gate $U_{\mathrm{ent}}=(Y\otimes Y)e^{-i(\pi/4)Z\otimes Z}$ between the memory qubit and the communication qubit up to an irrelevant global phase. Throughout this work, we use $X$, $Y$ and $Z$ to denote the three Pauli matrices.

As shown in Fig.~\ref{fig:scheme}\textbf{d}, after the ion-photon entanglement generation and the storage of the memory qubit, we execute different circuit modules based on the entangling gate $U_{\mathrm{ent}}$ and other single-qubit gates for different applications. In the experiment the photonic qubit is measured in the middle of the circuit to herald the successful establishment of ion-photon entanglement, but conceptually this measurement can be postponed to the end of the circuit without any change to the outcome. Finally we measure the states of the two ions by an electron multiplying charge-coupled device (EMCCD) using the electron shelving technique \cite{Roman2020,edmunds2020scalable,yang2022realizing}. Specifically, we transfer the population in $|0\rangle$ into $|0^\prime\rangle$ by global $411\,$nm and $3432\,$nm $\pi$ pulses, followed by another $411\,$nm pulse to shelve the residual population in $|0\rangle$ due to imperfect pulses into the $|{}^2 D_{5/2},F=2, m_F=-1\rangle$ Zeeman level. In this way, we suppress the detection error to be below $1\%$.

\subsection{Benchmarking fidelity of circuit segments}
\begin{figure*}[!tbp]
   \includegraphics[width=5.3in]{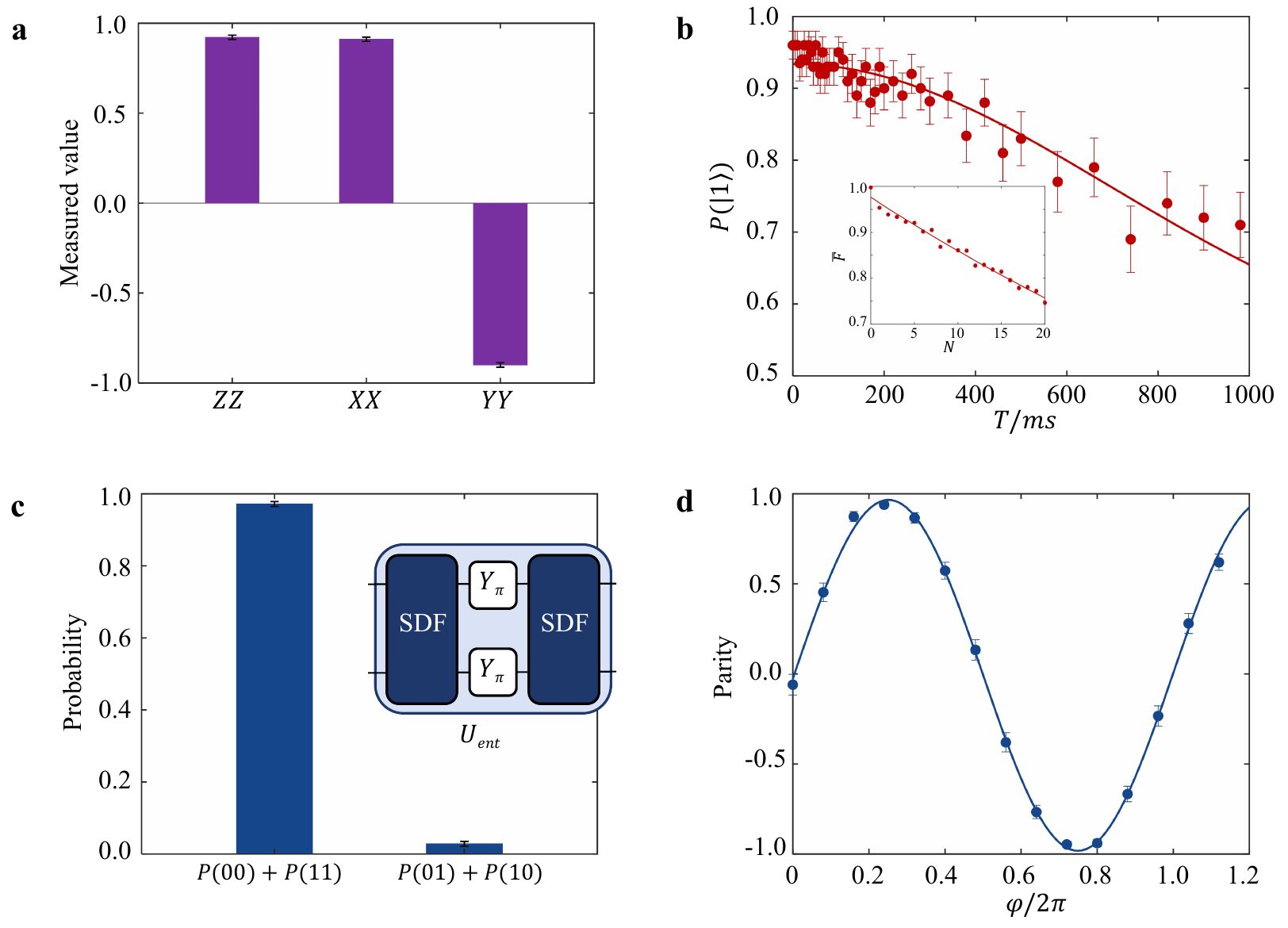}
   \caption{\textbf{Benchmarking elementary functions of the quantum network node.}
   \textbf{a}, Ion-photon correlation measurement gives an entanglement fidelity $F_{cp}=(1 + \langle ZZ\rangle + \langle XX \rangle - \langle YY \rangle)/4 = (93.3\pm0.5)\%$ between the communication qubit and the photonic qubit.
   \textbf{b}, We measure a Ramsey coherence time $T_2^* = (0.985 \pm 0.033)\,\mathrm{s}$ for the memory qubit without spin echoes. The inset shows the average storage fidelity $\bar{F}$ over six mutually unbiased basis states after $N$ rounds of $S$-$F$-$S$ qubit type conversions, from which a SPAM error $\epsilon_0=(2.4\pm0.4)\%$ and a round-trip conversion error $\epsilon=(1.26\pm0.04)\%$ can be fitted.
   \textbf{c, d}, The two-ion entangling gate between the memory qubit and the communication qubit. A Bell state fidelity of $F_{mc}=(96.3\pm0.8)\%$ is measured from the population in \textbf{c} and the parity in \textbf{d}. The entangling gate is realized by the time evolution under a spin-dependent force (SDF) with a global $\pi$ pulse in the middle, as shown in the inset. All the error bars represent one standard deviation.
   \label{fig:fidelity}}
\end{figure*}

First we calibrate the performance of individual components of our quantum network node. As shown in Fig.~\ref{fig:fidelity}\textbf{a}, we measure the communication qubit and the photonic qubit in different Pauli bases to obtain their correlations $\langle XX\rangle=0.91(1)$, $\langle YY\rangle=-0.90(1)$ and $\langle ZZ\rangle=0.92(1)$. Then the ion-photon entanglement fidelity is calculated to be $F_{cp}=(1 + \langle ZZ\rangle + \langle XX \rangle - \langle YY \rangle)/4 = (93.3\pm0.5)\%$. As discussed in our previous work \cite{Feng2024}, the infidelity comes from multiple sources including the state preparation and measure (SPAM) error, the dephasing of the Zeeman levels, the jitter time of the microwave pulses to combine the population in the Zeeman levels, the imperfect polarization of the $370\,$nm laser, and PMT dark counts. Compared with Ref.~\cite{Feng2024}, the entanglement fidelity is improved mainly because of the reduced detection error due to the use of electron shelving, the suppressed Zeeman dephasing error due to the shortened microwave pulse duration, and the decreased jitter time owing to the improvement in our electronic system. The shortened experimental sequence also leads to an increase in the entanglement generation rate to $r=7\,\mathrm{s}^{-1}$.

In Fig.~\ref{fig:fidelity}\textbf{b}, we perform a Ramsey experiment to measure the $T_2^*$ coherence time of the memory qubit. We initialize the memory qubit in $|+\rangle=(|0\rangle+|1\rangle)/\sqrt{2}$ in the $S$-type, convert it to the $F$-type for a storage time $T$, and finally convert it back for measurement.
Here we lock the microwave phases between the signal for the $S$-qubit microwave manipulation and that for the electro-optic modulator (EOM) which generates the two tones in the bichromatic $411\,$nm laser, so as to suppress the drift in the $S$-qubit basis over time.
In this way, we obtain a dephasing time $T_2^* = (0.985 \pm 0.033)\,$s without applying spin echoes.
Furthermore, at a storage time $T=50\,$ms, we measure an average memory fidelity $F_m=(93.7 \pm 0.1)\%$ over the six mutually unbiased basis (MUB) states \cite{WOOTTERS1989363}. The infidelity contains $\epsilon_0=(2.4\pm0.4)\%$ SPAM error and $\epsilon=(1.26\pm 0.04)\%$ round trip conversion error, which can be fitted by performing multiple rounds of qubit type conversions as shown in the inset, together with a memory error of $(1-e^{-T/T_2^*})/2= 2.5\%$.

To benchmark the fidelity of the entangling gate $U_{\mathrm{ent}}$, we use it to prepare a Bell state with the help of global microwave $\pi/2$ pulses \cite{PhysRevA.107.022617}. The population of $P_{00}+P_{11}$ is measured to be $(96.5\pm0.7)\%$ as shown in Fig.~\ref{fig:fidelity}\textbf{c}. By adding an analysis $\pi/2$ microwave pulse with a controllable phase $\varphi$ at the end of the sequence, we obtain the parity curve as shown in Fig.~\ref{fig:fidelity}\textbf{d} following $C\mathrm{sin}\varphi$ with a fitted contrast $|C|= (96.1\pm0.9)\%$. The Bell state fidelity is then calculated to be $F_{mc}=(96.3\pm0.8)\%$, with the error mainly coming from the SPAM error and the fluctuation of the trap frequency and the microwave intensity.

After calibrating these elementary components, we combine them for different applications of the quantum network node. In the following, we choose a fixed storage time $T=50\,$ms for the memory qubit in the experimental sequence in Fig.~\ref{fig:scheme}\textbf{d}, during which we attempt to generate the ion-photon entanglement on the communication qubit. Given the above $r=7\,\mathrm{s}^{-1}$ generation rate, we have $1-e^{-rT}=30\%$ probability to detect the photonic qubit. The data for the unsuccessful events are discarded.

\subsection{Quantum state teleportation}
\begin{figure}[!tbp]
   \includegraphics[width=3.3in]{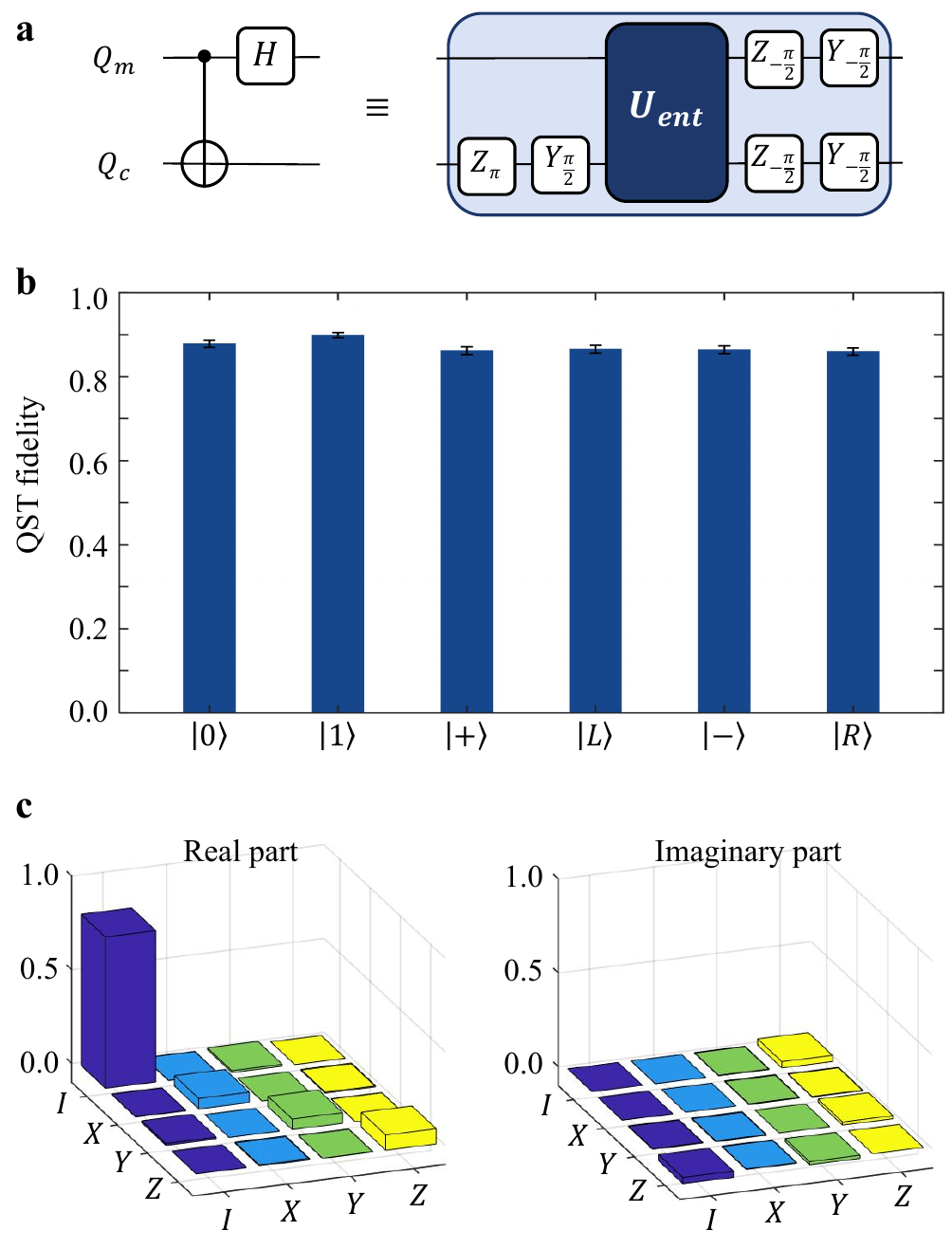}
   \caption{\textbf{Demonstration of quantum state teleportation.}
   \textbf{a}, The circuit module in Fig.~\ref{fig:scheme}\textbf{a} is decomposed into single-qubit rotations and a two-ion entangling gate to perform the Bell state measurement.
   \textbf{b}, Teleportation fidelities for the six  mutually unbiased basis states. Error bars represent one standard deviation.
   \textbf{c}, Real and imaginary parts of the reconstructed process matrix. A process fidelity $F_p=(80.7\pm0.6)\%$ is obtained.
   \label{fig:teleportation}}
\end{figure}

We use the quantum circuit in Fig.~\ref{fig:teleportation}\textbf{a} for Bell measurement as the circuit module in Fig.~\ref{fig:scheme}\textbf{d} to teleport the quantum state of the memory qubit to the photonic qubit. Specifically, we compile the CNOT gate and the Hadamard gate into our entangling gate $U_{\mathrm{ent}}$, single-qubit light shift gates $Z_{\theta}$ using an off-resonant global $411\,$nm laser beam, and single-qubit $Y_{\theta}$ rotations using a global microwave. To apply the single-qubit gates on the communication qubit alone, we place them before the qubit type conversion on the memory qubit, i.e., we hold the memory qubit in the $F$-type and apply the global pulses which solely affect the $S$-type communication qubit.

Now we measure the teleportation fidelity for the six MUB states. For each initial state of the memory qubit, we execute the experimental sequence and measure the photonic qubit in the $X$, $Y$ and $Z$ bases. Conceptually, the quantum teleportation requires a Pauli gate on the photonic qubit conditioned on the Bell measurement outcome of the two ionic qubits. However, here our photonic qubit has already been measured in the middle of the circuit to herald the successful ion-photon entanglement generation. Therefore, we interpret the required Pauli gate as a change in the measurement bases: a Pauli $X$ ($Y$, $Z$) gate will leave the measurement in the $X$ ($Y$, $Z$) basis unchanged, while flipping the sign of measurement outcomes in the other two bases. (See Supplementary Information for details.) In this way, we can recover the density matrix of the photonic qubit after the hypothetical conditional Pauli gate by the maximum likelihood method. For the initial state $|\psi\rangle_m= |0\rangle, |1\rangle, |+\rangle, |-\rangle, |L\rangle, |R\rangle$, we obtain the corresponding state fidelities $F=0.878(8), 0.899(7), 0.862(9), 0.866(9), 0.864(9), 0.860(9)$ as shown in Fig.~\ref{fig:teleportation}\textbf{b}, respectively, each with at least 1400 successful trials collected in each measurement basis. The average state fidelity is $\bar{F} = (87.2\pm0.9)\%$, exceeding the classical threshold of 2/3.
We further perform quantum process tomography \cite{MeasureTwoQubits} using the recovered density matrices of the six MUB states, and obtain the process matrix in Fig.~\ref{fig:teleportation}\textbf{c} by the maximum likelihood method. We obtain a process fidelity $F_p=(80.7\pm0.6)\%$, consistent with the average state fidelity $\bar{F}=(2F_P+1)/3$ \cite{PhysRevA.71.062310}. Here, error bars are computed as one standard deviation under Monte Carlo sampling, assuming a Poisson distribution of the photon counts.

\subsection{Three-qubit GHZ state}
\begin{figure}[!tbp]
   \includegraphics[width=2.68in]{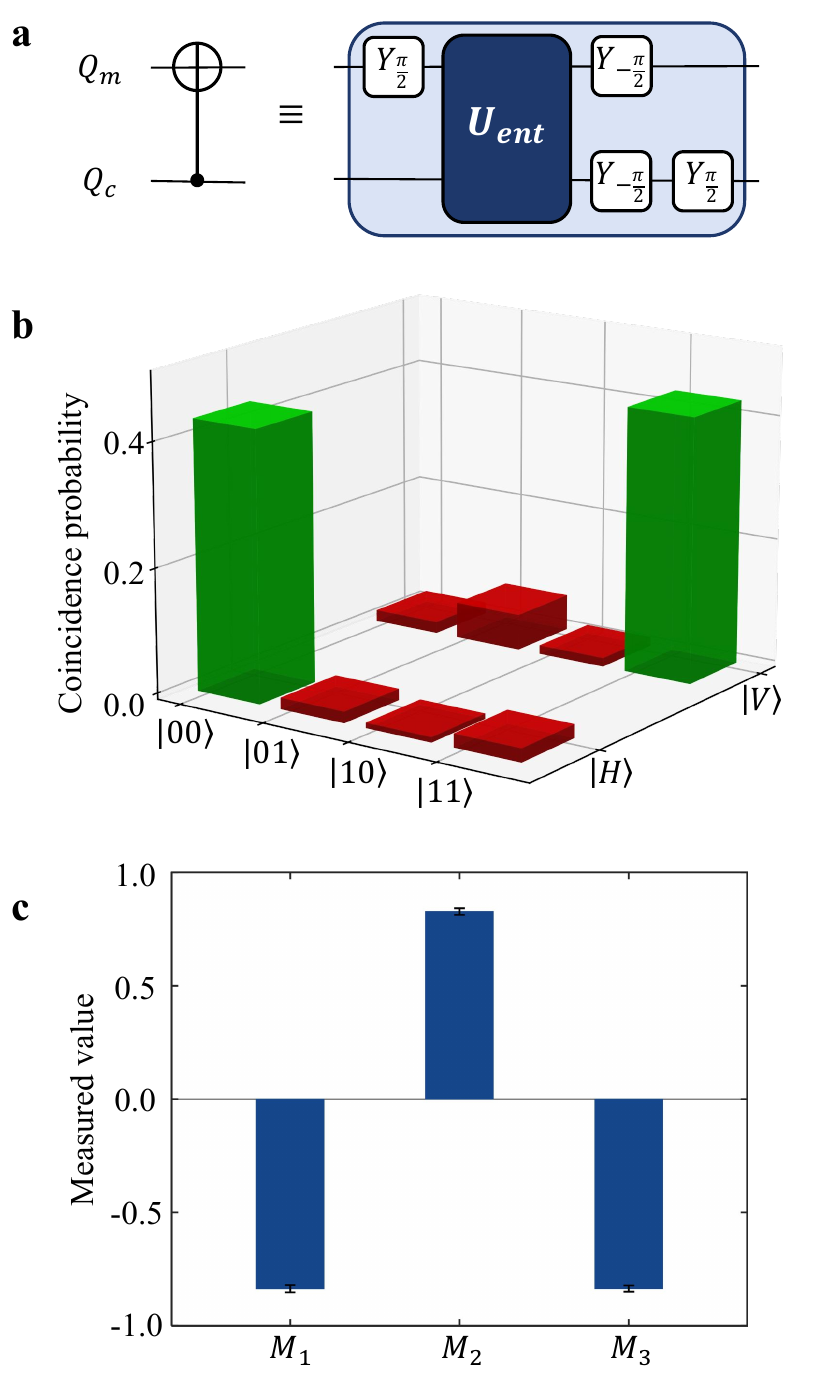}
   \caption{\textbf{Demonstration of a GHZ state preparation.} \textbf{a}, The circuit module in Fig.~\ref{fig:scheme}\textbf{a} is a CNOT gate and is decomposed into single-qubit rotations and a two-ion entangling gate.
   \textbf{b}, Normalized coincidence probability for ionic qubits in the $\{|0\rangle, |1\rangle\}$ basis and for the photonic qubit in the $\{|H\rangle, |V\rangle\}$ basis.
   \textbf{c}, Expectation values for $M_k^{\otimes 3}$ [$M_k=\mathrm{cos}(k\pi/3)X + \mathrm{sin}(k\pi/3)Y$, ($k = 1, 2, 3$)]. From them, a GHZ state fidelity $F=(84.7\pm 0.6)\%$ is deduced. Error bars represent one standard deviation.
   \label{fig:ghz}}
\end{figure}

We further use the quantum circuit in Fig.~\ref{fig:ghz}\textbf{a} as the circuit module in Fig.~\ref{fig:scheme}\textbf{d} to prepare a three-qubit GHZ state between the memory qubit, the communication qubit and the photonic qubit $|\mathrm{GHZ} \rangle_{mcp} = (|00H\rangle + |11V\rangle)/\sqrt{2}$. Here the first $Y_{\pi/2}$ rotation on the memory qubit is moved to the beginning of the circuit before the generation of ion-photon entanglement. After the two-ion entangling gate $U_{\mathrm{ent}}$, we apply a global microwave $Y_{-\pi/2}$ rotation with both ions in the $S$-type. Then we convert the memory qubit to the $F$-type again and use the global microwave for the individual $Y_{\pi/2}$ gate on the communication qubit.

The fidelity of the GHZ state can be measured by decomposing into local observables \cite{PhysRevA.76.030305}:
\begin{align}
|\mathrm{GHZ}_3\rangle \langle \mathrm{GHZ}_3| = & \frac{1}{2}(|00H\rangle \langle 00H| +|11V\rangle \langle 11V| ) \nonumber\\
& + \frac{1}{6}\sum_{k=1}^3(-1)^k M_k^{\otimes 3},
\label{eq:witness}
\end{align}
where $M_k=\mathrm{cos}(k\pi/3)X + \mathrm{sin}(k\pi/3)Y$ ($k = 1, 2, 3$). For the first item, we measure the two ions in the $Z$ basis and the photon in the $|H\rangle/|V\rangle$ basis to get the population distribution in Fig.~\ref{fig:ghz}\textbf{b}, and obtain $P_{00H}=0.434(11)$ and $P_{11V}=0.427(11)$ with 2200 successful trials. As for the second term, we apply a global microwave $\pi/2$ pulse to rotate the two ions into the $M_k$ basis, and set the angles of the waveplates for the measurement basis of the photon (see Supplementary Information for details). For each $k$, we collect 1200 successful data points, and obtain $M_1^{\otimes 3}=-0.84(2)$, $M_2^{\otimes 3}=0.82(2)$ and $M_3^{\otimes 3}=-0.84(1)$ as shown in Fig.~\ref{fig:ghz}\textbf{c}. Combining these results, we calculate the GHZ state fidelity as $F=(84.7\pm 0.6)\%$. Again the error bar represents one standard deviation by assuming a Poisson distribution of the photon counts.

\section{Discussion}
Our measured teleportation fidelity and GHZ state fidelity are consistent with the concatenation of the errors from the three elementary components $F_m F_{cp} F_{mc}\approx 84\%$, with the GHZ state fidelity being slightly lower than that of quantum state teleportation due to one more rounds of qubit type conversion. This suggests that there is no significant coherent error in our system, and that the calibrated elementary operations are stable over the whole execution time. To further enhance the fidelities in the future, we may optimize the electronic system to shorten the jitter time, and suppress the detection error by shelving the $|0\rangle$ state into multiple Zeeman levels of ${}^2F_{7/2}$ \cite{Roman2020,edmunds2020scalable,yang2022realizing}. We may also add active stabilization of the microwave intensity to suppress its slow drift.

In summary, we have realized a fully-functioning dual-type trapped-ion quantum network node, with the capability of crosstalk-free ion-photon entanglement generation and quantum state storage, and entangling gates between the communication qubit and the memory qubit. We demonstrate its applications like the quantum state teleportation and the preparation of multipartite entangled states. By further connecting multiple nodes via photon interference, and by further increasing the ion number in the trap, a larger-scale quantum network can be realized for its wide applications in quantum cryptography \cite{Gisin2007,RevModPhys.81.1301,RevModPhys.92.025002}, quantum metrology \cite{doi:10.1126/science.1104149,Komar2014,PhysRevLett.109.070503} and quantum computing \cite{PhysRevA.59.4249,PhysRevA.76.062323,broadbent2009blind}.

\textbf{Data Availability:} The data that support the findings of this study are available
from the authors upon request.

\textbf{Acknowledgements:} This work was supported by Innovation Program for Quantum Science and Technology (grant nos. 2021ZD0301601 and 2021ZD0301604), Tsinghua University Initiative Scientific Research Program, and the Ministry of Education of China. L.M.D. acknowledges in addition support from the New Cornerstone Science Foundation through the New Cornerstone Investigator Program. Y.K.W. and Y.F.P. acknowledge in addition support from the Dushi program from Tsinghua University.

\textbf{Competing interests:} C.Z. and Q.X.M. are affiliated with HYQ Co. Y.Y.H, L.F, Y.K.W., B.X.Q., Y.X., Y.F.P., Z.C.Z. and L.M.D. hold shares with HYQ Co. The other authors declare no competing interests.

\textbf{Author Information:} Correspondence and requests for materials should be addressed to L.M.D.
(lmduan@tsinghua.edu.cn).

\textbf{Author Contributions:} Y.Y.H., L.F., Y.K.W., Y.L.X., L.Z., Z.B.C, C.X.H., C.Z., S.A.G., Q.X.M., B.X.Q., Y.X., Y.F.P, and Z.C.Z. carried out the experiment and analyzed the data. L.M.D. proposed the experiment and supervised the project. Y.K.W.did the associated theoretical analysis. Y.K.W., Y.Y.H., L.F and L.M.D. wrote the manuscript.


\begin{thebibliography}{53}%
\makeatletter
\providecommand \@ifxundefined [1]{%
 \@ifx{#1\undefined}
}%
\providecommand \@ifnum [1]{%
 \ifnum #1\expandafter \@firstoftwo
 \else \expandafter \@secondoftwo
 \fi
}%
\providecommand \@ifx [1]{%
 \ifx #1\expandafter \@firstoftwo
 \else \expandafter \@secondoftwo
 \fi
}%
\providecommand \natexlab [1]{#1}%
\providecommand \enquote  [1]{``#1''}%
\providecommand \bibnamefont  [1]{#1}%
\providecommand \bibfnamefont [1]{#1}%
\providecommand \citenamefont [1]{#1}%
\providecommand \href@noop [0]{\@secondoftwo}%
\providecommand \href [0]{\begingroup \@sanitize@url \@href}%
\providecommand \@href[1]{\@@startlink{#1}\@@href}%
\providecommand \@@href[1]{\endgroup#1\@@endlink}%
\providecommand \@sanitize@url [0]{\catcode `\\12\catcode `\$12\catcode
  `\&12\catcode `\#12\catcode `\^12\catcode `\_12\catcode `\%12\relax}%
\providecommand \@@startlink[1]{}%
\providecommand \@@endlink[0]{}%
\providecommand \url  [0]{\begingroup\@sanitize@url \@url }%
\providecommand \@url [1]{\endgroup\@href {#1}{\urlprefix }}%
\providecommand \urlprefix  [0]{URL }%
\providecommand \Eprint [0]{\href }%
\providecommand \doibase [0]{http://dx.doi.org/}%
\providecommand \selectlanguage [0]{\@gobble}%
\providecommand \bibinfo  [0]{\@secondoftwo}%
\providecommand \bibfield  [0]{\@secondoftwo}%
\providecommand \translation [1]{[#1]}%
\providecommand \BibitemOpen [0]{}%
\providecommand \bibitemStop [0]{}%
\providecommand \bibitemNoStop [0]{.\EOS\space}%
\providecommand \EOS [0]{\spacefactor3000\relax}%
\providecommand \BibitemShut  [1]{\csname bibitem#1\endcsname}%
\let\auto@bib@innerbib\@empty
\bibitem [{\citenamefont {Kimble}(2008)}]{Kimble2008}%
  \BibitemOpen
  \bibfield  {author} {\bibinfo {author} {\bibfnamefont {H.~J.}\ \bibnamefont
  {Kimble}},\ }\bibfield  {title} {\enquote {\bibinfo {title} {The quantum
  internet},}\ }\href {\doibase 10.1038/nature07127} {\bibfield  {journal}
  {\bibinfo  {journal} {Nature}\ }\textbf {\bibinfo {volume} {453}},\ \bibinfo
  {pages} {1023--1030} (\bibinfo {year} {2008})}\BibitemShut {NoStop}%
\bibitem [{\citenamefont {Wehner}\ \emph {et~al.}(2018)\citenamefont {Wehner},
  \citenamefont {Elkouss},\ and\ \citenamefont
  {Hanson}}]{doi:10.1126/science.aam9288}%
  \BibitemOpen
  \bibfield  {author} {\bibinfo {author} {\bibfnamefont {Stephanie}\
  \bibnamefont {Wehner}}, \bibinfo {author} {\bibfnamefont {David}\
  \bibnamefont {Elkouss}}, \ and\ \bibinfo {author} {\bibfnamefont {Ronald}\
  \bibnamefont {Hanson}},\ }\bibfield  {title} {\enquote {\bibinfo {title}
  {Quantum internet: A vision for the road ahead},}\ }\href {\doibase
  10.1126/science.aam9288} {\bibfield  {journal} {\bibinfo  {journal}
  {Science}\ }\textbf {\bibinfo {volume} {362}},\ \bibinfo {pages} {eaam9288}
  (\bibinfo {year} {2018})}\BibitemShut {NoStop}%
\bibitem [{\citenamefont {Gisin}\ and\ \citenamefont {Thew}(2007)}]{Gisin2007}%
  \BibitemOpen
  \bibfield  {author} {\bibinfo {author} {\bibfnamefont {Nicolas}\ \bibnamefont
  {Gisin}}\ and\ \bibinfo {author} {\bibfnamefont {Rob}\ \bibnamefont {Thew}},\
  }\bibfield  {title} {\enquote {\bibinfo {title} {Quantum communication},}\
  }\href {\doibase 10.1038/nphoton.2007.22} {\bibfield  {journal} {\bibinfo
  {journal} {Nature Photonics}\ }\textbf {\bibinfo {volume} {1}},\ \bibinfo
  {pages} {165--171} (\bibinfo {year} {2007})}\BibitemShut {NoStop}%
\bibitem [{\citenamefont {Scarani}\ \emph {et~al.}(2009)\citenamefont
  {Scarani}, \citenamefont {Bechmann-Pasquinucci}, \citenamefont {Cerf},
  \citenamefont {Du\ifmmode~\check{s}\else \v{s}\fi{}ek}, \citenamefont
  {L\"utkenhaus},\ and\ \citenamefont {Peev}}]{RevModPhys.81.1301}%
  \BibitemOpen
  \bibfield  {author} {\bibinfo {author} {\bibfnamefont {Valerio}\ \bibnamefont
  {Scarani}}, \bibinfo {author} {\bibfnamefont {Helle}\ \bibnamefont
  {Bechmann-Pasquinucci}}, \bibinfo {author} {\bibfnamefont {Nicolas~J.}\
  \bibnamefont {Cerf}}, \bibinfo {author} {\bibfnamefont {Miloslav}\
  \bibnamefont {Du\ifmmode~\check{s}\else \v{s}\fi{}ek}}, \bibinfo {author}
  {\bibfnamefont {Norbert}\ \bibnamefont {L\"utkenhaus}}, \ and\ \bibinfo
  {author} {\bibfnamefont {Momtchil}\ \bibnamefont {Peev}},\ }\bibfield
  {title} {\enquote {\bibinfo {title} {The security of practical quantum key
  distribution},}\ }\href {\doibase 10.1103/RevModPhys.81.1301} {\bibfield
  {journal} {\bibinfo  {journal} {Rev. Mod. Phys.}\ }\textbf {\bibinfo {volume}
  {81}},\ \bibinfo {pages} {1301--1350} (\bibinfo {year} {2009})}\BibitemShut
  {NoStop}%
\bibitem [{\citenamefont {Xu}\ \emph {et~al.}(2020)\citenamefont {Xu},
  \citenamefont {Ma}, \citenamefont {Zhang}, \citenamefont {Lo},\ and\
  \citenamefont {Pan}}]{RevModPhys.92.025002}%
  \BibitemOpen
  \bibfield  {author} {\bibinfo {author} {\bibfnamefont {Feihu}\ \bibnamefont
  {Xu}}, \bibinfo {author} {\bibfnamefont {Xiongfeng}\ \bibnamefont {Ma}},
  \bibinfo {author} {\bibfnamefont {Qiang}\ \bibnamefont {Zhang}}, \bibinfo
  {author} {\bibfnamefont {Hoi-Kwong}\ \bibnamefont {Lo}}, \ and\ \bibinfo
  {author} {\bibfnamefont {Jian-Wei}\ \bibnamefont {Pan}},\ }\bibfield  {title}
  {\enquote {\bibinfo {title} {Secure quantum key distribution with realistic
  devices},}\ }\href {\doibase 10.1103/RevModPhys.92.025002} {\bibfield
  {journal} {\bibinfo  {journal} {Rev. Mod. Phys.}\ }\textbf {\bibinfo {volume}
  {92}},\ \bibinfo {pages} {025002} (\bibinfo {year} {2020})}\BibitemShut
  {NoStop}%
\bibitem [{\citenamefont {Giovannetti}\ \emph {et~al.}(2004)\citenamefont
  {Giovannetti}, \citenamefont {Lloyd},\ and\ \citenamefont
  {Maccone}}]{doi:10.1126/science.1104149}%
  \BibitemOpen
  \bibfield  {author} {\bibinfo {author} {\bibfnamefont {Vittorio}\
  \bibnamefont {Giovannetti}}, \bibinfo {author} {\bibfnamefont {Seth}\
  \bibnamefont {Lloyd}}, \ and\ \bibinfo {author} {\bibfnamefont {Lorenzo}\
  \bibnamefont {Maccone}},\ }\bibfield  {title} {\enquote {\bibinfo {title}
  {Quantum-enhanced measurements: Beating the standard quantum limit},}\ }\href
  {\doibase 10.1126/science.1104149} {\bibfield  {journal} {\bibinfo  {journal}
  {Science}\ }\textbf {\bibinfo {volume} {306}},\ \bibinfo {pages} {1330--1336}
  (\bibinfo {year} {2004})}\BibitemShut {NoStop}%
\bibitem [{\citenamefont {K{\'o}m{\'a}r}\ \emph {et~al.}(2014)\citenamefont
  {K{\'o}m{\'a}r}, \citenamefont {Kessler}, \citenamefont {Bishof},
  \citenamefont {Jiang}, \citenamefont {S{\o}rensen}, \citenamefont {Ye},\ and\
  \citenamefont {Lukin}}]{Komar2014}%
  \BibitemOpen
  \bibfield  {author} {\bibinfo {author} {\bibfnamefont {P.}~\bibnamefont
  {K{\'o}m{\'a}r}}, \bibinfo {author} {\bibfnamefont {E.~M.}\ \bibnamefont
  {Kessler}}, \bibinfo {author} {\bibfnamefont {M.}~\bibnamefont {Bishof}},
  \bibinfo {author} {\bibfnamefont {L.}~\bibnamefont {Jiang}}, \bibinfo
  {author} {\bibfnamefont {A.~S.}\ \bibnamefont {S{\o}rensen}}, \bibinfo
  {author} {\bibfnamefont {J.}~\bibnamefont {Ye}}, \ and\ \bibinfo {author}
  {\bibfnamefont {M.~D.}\ \bibnamefont {Lukin}},\ }\bibfield  {title} {\enquote
  {\bibinfo {title} {A quantum network of clocks},}\ }\href {\doibase
  10.1038/nphys3000} {\bibfield  {journal} {\bibinfo  {journal} {Nature
  Physics}\ }\textbf {\bibinfo {volume} {10}},\ \bibinfo {pages} {582--587}
  (\bibinfo {year} {2014})}\BibitemShut {NoStop}%
\bibitem [{\citenamefont {Gottesman}\ \emph {et~al.}(2012)\citenamefont
  {Gottesman}, \citenamefont {Jennewein},\ and\ \citenamefont
  {Croke}}]{PhysRevLett.109.070503}%
  \BibitemOpen
  \bibfield  {author} {\bibinfo {author} {\bibfnamefont {Daniel}\ \bibnamefont
  {Gottesman}}, \bibinfo {author} {\bibfnamefont {Thomas}\ \bibnamefont
  {Jennewein}}, \ and\ \bibinfo {author} {\bibfnamefont {Sarah}\ \bibnamefont
  {Croke}},\ }\bibfield  {title} {\enquote {\bibinfo {title} {Longer-baseline
  telescopes using quantum repeaters},}\ }\href {\doibase
  10.1103/PhysRevLett.109.070503} {\bibfield  {journal} {\bibinfo  {journal}
  {Phys. Rev. Lett.}\ }\textbf {\bibinfo {volume} {109}},\ \bibinfo {pages}
  {070503} (\bibinfo {year} {2012})}\BibitemShut {NoStop}%
\bibitem [{\citenamefont {Cirac}\ \emph {et~al.}(1999)\citenamefont {Cirac},
  \citenamefont {Ekert}, \citenamefont {Huelga},\ and\ \citenamefont
  {Macchiavello}}]{PhysRevA.59.4249}%
  \BibitemOpen
  \bibfield  {author} {\bibinfo {author} {\bibfnamefont {J.~I.}\ \bibnamefont
  {Cirac}}, \bibinfo {author} {\bibfnamefont {A.~K.}\ \bibnamefont {Ekert}},
  \bibinfo {author} {\bibfnamefont {S.~F.}\ \bibnamefont {Huelga}}, \ and\
  \bibinfo {author} {\bibfnamefont {C.}~\bibnamefont {Macchiavello}},\
  }\bibfield  {title} {\enquote {\bibinfo {title} {Distributed quantum
  computation over noisy channels},}\ }\href {\doibase
  10.1103/PhysRevA.59.4249} {\bibfield  {journal} {\bibinfo  {journal} {Phys.
  Rev. A}\ }\textbf {\bibinfo {volume} {59}},\ \bibinfo {pages} {4249--4254}
  (\bibinfo {year} {1999})}\BibitemShut {NoStop}%
\bibitem [{\citenamefont {Jiang}\ \emph {et~al.}(2007)\citenamefont {Jiang},
  \citenamefont {Taylor}, \citenamefont {S\o{}rensen},\ and\ \citenamefont
  {Lukin}}]{PhysRevA.76.062323}%
  \BibitemOpen
  \bibfield  {author} {\bibinfo {author} {\bibfnamefont {Liang}\ \bibnamefont
  {Jiang}}, \bibinfo {author} {\bibfnamefont {Jacob~M.}\ \bibnamefont
  {Taylor}}, \bibinfo {author} {\bibfnamefont {Anders~S.}\ \bibnamefont
  {S\o{}rensen}}, \ and\ \bibinfo {author} {\bibfnamefont {Mikhail~D.}\
  \bibnamefont {Lukin}},\ }\bibfield  {title} {\enquote {\bibinfo {title}
  {Distributed quantum computation based on small quantum registers},}\ }\href
  {\doibase 10.1103/PhysRevA.76.062323} {\bibfield  {journal} {\bibinfo
  {journal} {Phys. Rev. A}\ }\textbf {\bibinfo {volume} {76}},\ \bibinfo
  {pages} {062323} (\bibinfo {year} {2007})}\BibitemShut {NoStop}%
\bibitem [{\citenamefont {Broadbent}\ \emph {et~al.}(2009)\citenamefont
  {Broadbent}, \citenamefont {Fitzsimons},\ and\ \citenamefont
  {Kashefi}}]{broadbent2009blind}%
  \BibitemOpen
  \bibfield  {author} {\bibinfo {author} {\bibfnamefont {Anne}\ \bibnamefont
  {Broadbent}}, \bibinfo {author} {\bibfnamefont {Joseph}\ \bibnamefont
  {Fitzsimons}}, \ and\ \bibinfo {author} {\bibfnamefont {Elham}\ \bibnamefont
  {Kashefi}},\ }\bibfield  {title} {\enquote {\bibinfo {title} {Universal blind
  quantum computation},}\ }in\ \href {\doibase 10.1109/FOCS.2009.36} {\emph
  {\bibinfo {booktitle} {2009 50th Annual IEEE Symposium on Foundations of
  Computer Science}}}\ (\bibinfo {year} {2009})\ pp.\ \bibinfo {pages}
  {517--526}\BibitemShut {NoStop}%
\bibitem [{\citenamefont {Duan}\ \emph {et~al.}(2004)\citenamefont {Duan},
  \citenamefont {Blinov}, \citenamefont {Moehring},\ and\ \citenamefont
  {Monroe}}]{10.5555/2011617.2011618}%
  \BibitemOpen
  \bibfield  {author} {\bibinfo {author} {\bibfnamefont {L.-M.}\ \bibnamefont
  {Duan}}, \bibinfo {author} {\bibfnamefont {B.~B.}\ \bibnamefont {Blinov}},
  \bibinfo {author} {\bibfnamefont {D.~L.}\ \bibnamefont {Moehring}}, \ and\
  \bibinfo {author} {\bibfnamefont {C.}~\bibnamefont {Monroe}},\ }\bibfield
  {title} {\enquote {\bibinfo {title} {Scalable trapped ion quantum computation
  with a probabilistic ion-photon mapping},}\ }\href@noop {} {\bibfield
  {journal} {\bibinfo  {journal} {Quantum Info. Comput.}\ }\textbf {\bibinfo
  {volume} {4}},\ \bibinfo {pages} {165--173} (\bibinfo {year}
  {2004})}\BibitemShut {NoStop}%
\bibitem [{\citenamefont {Duan}\ and\ \citenamefont
  {Monroe}(2010)}]{RevModPhys.82.1209}%
  \BibitemOpen
  \bibfield  {author} {\bibinfo {author} {\bibfnamefont {L.-M.}\ \bibnamefont
  {Duan}}\ and\ \bibinfo {author} {\bibfnamefont {C.}~\bibnamefont {Monroe}},\
  }\bibfield  {title} {\enquote {\bibinfo {title} {Colloquium: Quantum networks
  with trapped ions},}\ }\href {\doibase 10.1103/RevModPhys.82.1209} {\bibfield
   {journal} {\bibinfo  {journal} {Rev. Mod. Phys.}\ }\textbf {\bibinfo
  {volume} {82}},\ \bibinfo {pages} {1209--1224} (\bibinfo {year}
  {2010})}\BibitemShut {NoStop}%
\bibitem [{\citenamefont {Duan}\ \emph {et~al.}(2001)\citenamefont {Duan},
  \citenamefont {Lukin}, \citenamefont {Cirac},\ and\ \citenamefont
  {Zoller}}]{Duan2001DLCZ}%
  \BibitemOpen
  \bibfield  {author} {\bibinfo {author} {\bibfnamefont {L.-M.}\ \bibnamefont
  {Duan}}, \bibinfo {author} {\bibfnamefont {M.~D.}\ \bibnamefont {Lukin}},
  \bibinfo {author} {\bibfnamefont {J.~I.}\ \bibnamefont {Cirac}}, \ and\
  \bibinfo {author} {\bibfnamefont {P.}~\bibnamefont {Zoller}},\ }\bibfield
  {title} {\enquote {\bibinfo {title} {Long-distance quantum communication with
  atomic ensembles and linear optics},}\ }\href {\doibase 10.1038/35106500}
  {\bibfield  {journal} {\bibinfo  {journal} {Nature}\ }\textbf {\bibinfo
  {volume} {414}},\ \bibinfo {pages} {413--418} (\bibinfo {year}
  {2001})}\BibitemShut {NoStop}%
\bibitem [{\citenamefont {Hammerer}\ \emph {et~al.}(2010)\citenamefont
  {Hammerer}, \citenamefont {S\o{}rensen},\ and\ \citenamefont
  {Polzik}}]{RevModPhys.82.1041}%
  \BibitemOpen
  \bibfield  {author} {\bibinfo {author} {\bibfnamefont {Klemens}\ \bibnamefont
  {Hammerer}}, \bibinfo {author} {\bibfnamefont {Anders~S.}\ \bibnamefont
  {S\o{}rensen}}, \ and\ \bibinfo {author} {\bibfnamefont {Eugene~S.}\
  \bibnamefont {Polzik}},\ }\bibfield  {title} {\enquote {\bibinfo {title}
  {Quantum interface between light and atomic ensembles},}\ }\href {\doibase
  10.1103/RevModPhys.82.1041} {\bibfield  {journal} {\bibinfo  {journal} {Rev.
  Mod. Phys.}\ }\textbf {\bibinfo {volume} {82}},\ \bibinfo {pages}
  {1041--1093} (\bibinfo {year} {2010})}\BibitemShut {NoStop}%
\bibitem [{\citenamefont {Sangouard}\ \emph {et~al.}(2011)\citenamefont
  {Sangouard}, \citenamefont {Simon}, \citenamefont {de~Riedmatten},\ and\
  \citenamefont {Gisin}}]{RevModPhys.83.33}%
  \BibitemOpen
  \bibfield  {author} {\bibinfo {author} {\bibfnamefont {Nicolas}\ \bibnamefont
  {Sangouard}}, \bibinfo {author} {\bibfnamefont {Christoph}\ \bibnamefont
  {Simon}}, \bibinfo {author} {\bibfnamefont {Hugues}\ \bibnamefont
  {de~Riedmatten}}, \ and\ \bibinfo {author} {\bibfnamefont {Nicolas}\
  \bibnamefont {Gisin}},\ }\bibfield  {title} {\enquote {\bibinfo {title}
  {Quantum repeaters based on atomic ensembles and linear optics},}\ }\href
  {\doibase 10.1103/RevModPhys.83.33} {\bibfield  {journal} {\bibinfo
  {journal} {Rev. Mod. Phys.}\ }\textbf {\bibinfo {volume} {83}},\ \bibinfo
  {pages} {33--80} (\bibinfo {year} {2011})}\BibitemShut {NoStop}%
\bibitem [{\citenamefont {Reiserer}\ and\ \citenamefont
  {Rempe}(2015)}]{RevModPhys.87.1379}%
  \BibitemOpen
  \bibfield  {author} {\bibinfo {author} {\bibfnamefont {Andreas}\ \bibnamefont
  {Reiserer}}\ and\ \bibinfo {author} {\bibfnamefont {Gerhard}\ \bibnamefont
  {Rempe}},\ }\bibfield  {title} {\enquote {\bibinfo {title} {Cavity-based
  quantum networks with single atoms and optical photons},}\ }\href {\doibase
  10.1103/RevModPhys.87.1379} {\bibfield  {journal} {\bibinfo  {journal} {Rev.
  Mod. Phys.}\ }\textbf {\bibinfo {volume} {87}},\ \bibinfo {pages}
  {1379--1418} (\bibinfo {year} {2015})}\BibitemShut {NoStop}%
\bibitem [{\citenamefont {Reiserer}(2022)}]{RevModPhys.94.041003}%
  \BibitemOpen
  \bibfield  {author} {\bibinfo {author} {\bibfnamefont {Andreas}\ \bibnamefont
  {Reiserer}},\ }\bibfield  {title} {\enquote {\bibinfo {title} {Colloquium:
  Cavity-enhanced quantum network nodes},}\ }\href {\doibase
  10.1103/RevModPhys.94.041003} {\bibfield  {journal} {\bibinfo  {journal}
  {Rev. Mod. Phys.}\ }\textbf {\bibinfo {volume} {94}},\ \bibinfo {pages}
  {041003} (\bibinfo {year} {2022})}\BibitemShut {NoStop}%
\bibitem [{\citenamefont {Childress}\ and\ \citenamefont
  {Hanson}(2013)}]{Childress_Hanson_2013}%
  \BibitemOpen
  \bibfield  {author} {\bibinfo {author} {\bibfnamefont {Lilian}\ \bibnamefont
  {Childress}}\ and\ \bibinfo {author} {\bibfnamefont {Ronald}\ \bibnamefont
  {Hanson}},\ }\bibfield  {title} {\enquote {\bibinfo {title} {Diamond nv
  centers for quantum computing and quantum networks},}\ }\href {\doibase
  10.1557/mrs.2013.20} {\bibfield  {journal} {\bibinfo  {journal} {MRS
  Bulletin}\ }\textbf {\bibinfo {volume} {38}},\ \bibinfo {pages} {134–138}
  (\bibinfo {year} {2013})}\BibitemShut {NoStop}%
\bibitem [{\citenamefont {Awschalom}\ \emph {et~al.}(2018)\citenamefont
  {Awschalom}, \citenamefont {Hanson}, \citenamefont {Wrachtrup},\ and\
  \citenamefont {Zhou}}]{Awschalom2018}%
  \BibitemOpen
  \bibfield  {author} {\bibinfo {author} {\bibfnamefont {David~D.}\
  \bibnamefont {Awschalom}}, \bibinfo {author} {\bibfnamefont {Ronald}\
  \bibnamefont {Hanson}}, \bibinfo {author} {\bibfnamefont {J{\"o}rg}\
  \bibnamefont {Wrachtrup}}, \ and\ \bibinfo {author} {\bibfnamefont
  {Brian~B.}\ \bibnamefont {Zhou}},\ }\bibfield  {title} {\enquote {\bibinfo
  {title} {Quantum technologies with optically interfaced solid-state spins},}\
  }\href {\doibase 10.1038/s41566-018-0232-2} {\bibfield  {journal} {\bibinfo
  {journal} {Nature Photonics}\ }\textbf {\bibinfo {volume} {12}},\ \bibinfo
  {pages} {516--527} (\bibinfo {year} {2018})}\BibitemShut {NoStop}%
\bibitem [{\citenamefont {Mirhosseini}\ \emph {et~al.}(2020)\citenamefont
  {Mirhosseini}, \citenamefont {Sipahigil}, \citenamefont {Kalaee},\ and\
  \citenamefont {Painter}}]{Mirhosseini2020}%
  \BibitemOpen
  \bibfield  {author} {\bibinfo {author} {\bibfnamefont {Mohammad}\
  \bibnamefont {Mirhosseini}}, \bibinfo {author} {\bibfnamefont {Alp}\
  \bibnamefont {Sipahigil}}, \bibinfo {author} {\bibfnamefont {Mahmoud}\
  \bibnamefont {Kalaee}}, \ and\ \bibinfo {author} {\bibfnamefont {Oskar}\
  \bibnamefont {Painter}},\ }\bibfield  {title} {\enquote {\bibinfo {title}
  {Superconducting qubit to optical photon transduction},}\ }\href {\doibase
  10.1038/s41586-020-3038-6} {\bibfield  {journal} {\bibinfo  {journal}
  {Nature}\ }\textbf {\bibinfo {volume} {588}},\ \bibinfo {pages} {599--603}
  (\bibinfo {year} {2020})}\BibitemShut {NoStop}%
\bibitem [{\citenamefont {Zhong}\ \emph {et~al.}(2021)\citenamefont {Zhong},
  \citenamefont {Chang}, \citenamefont {Bienfait}, \citenamefont {Dumur},
  \citenamefont {Chou}, \citenamefont {Conner}, \citenamefont {Grebel},
  \citenamefont {Povey}, \citenamefont {Yan}, \citenamefont {Schuster},\ and\
  \citenamefont {Cleland}}]{Zhong2021}%
  \BibitemOpen
  \bibfield  {author} {\bibinfo {author} {\bibfnamefont {Youpeng}\ \bibnamefont
  {Zhong}}, \bibinfo {author} {\bibfnamefont {Hung-Shen}\ \bibnamefont
  {Chang}}, \bibinfo {author} {\bibfnamefont {Audrey}\ \bibnamefont
  {Bienfait}}, \bibinfo {author} {\bibfnamefont {{\'E}tienne}\ \bibnamefont
  {Dumur}}, \bibinfo {author} {\bibfnamefont {Ming-Han}\ \bibnamefont {Chou}},
  \bibinfo {author} {\bibfnamefont {Christopher~R.}\ \bibnamefont {Conner}},
  \bibinfo {author} {\bibfnamefont {Joel}\ \bibnamefont {Grebel}}, \bibinfo
  {author} {\bibfnamefont {Rhys~G.}\ \bibnamefont {Povey}}, \bibinfo {author}
  {\bibfnamefont {Haoxiong}\ \bibnamefont {Yan}}, \bibinfo {author}
  {\bibfnamefont {David~I.}\ \bibnamefont {Schuster}}, \ and\ \bibinfo {author}
  {\bibfnamefont {Andrew~N.}\ \bibnamefont {Cleland}},\ }\bibfield  {title}
  {\enquote {\bibinfo {title} {Deterministic multi-qubit entanglement in a
  quantum network},}\ }\href {\doibase 10.1038/s41586-021-03288-7} {\bibfield
  {journal} {\bibinfo  {journal} {Nature}\ }\textbf {\bibinfo {volume} {590}},\
  \bibinfo {pages} {571--575} (\bibinfo {year} {2021})}\BibitemShut {NoStop}%
\bibitem [{\citenamefont {Wang}\ \emph {et~al.}(2021)\citenamefont {Wang},
  \citenamefont {Luan}, \citenamefont {Qiao}, \citenamefont {Um}, \citenamefont
  {Zhang}, \citenamefont {Wang}, \citenamefont {Yuan}, \citenamefont {Gu},
  \citenamefont {Zhang},\ and\ \citenamefont {Kim}}]{wang2021single}%
  \BibitemOpen
  \bibfield  {author} {\bibinfo {author} {\bibfnamefont {Pengfei}\ \bibnamefont
  {Wang}}, \bibinfo {author} {\bibfnamefont {Chun-Yang}\ \bibnamefont {Luan}},
  \bibinfo {author} {\bibfnamefont {Mu}~\bibnamefont {Qiao}}, \bibinfo {author}
  {\bibfnamefont {Mark}\ \bibnamefont {Um}}, \bibinfo {author} {\bibfnamefont
  {Junhua}\ \bibnamefont {Zhang}}, \bibinfo {author} {\bibfnamefont
  {Ye}~\bibnamefont {Wang}}, \bibinfo {author} {\bibfnamefont {Xiao}\
  \bibnamefont {Yuan}}, \bibinfo {author} {\bibfnamefont {Mile}\ \bibnamefont
  {Gu}}, \bibinfo {author} {\bibfnamefont {Jingning}\ \bibnamefont {Zhang}}, \
  and\ \bibinfo {author} {\bibfnamefont {Kihwan}\ \bibnamefont {Kim}},\
  }\bibfield  {title} {\enquote {\bibinfo {title} {Single ion qubit with
  estimated coherence time exceeding one hour},}\ }\href
  {https://doi.org/10.1038/s41467-020-20330-w} {\bibfield  {journal} {\bibinfo
  {journal} {Nature communications}\ }\textbf {\bibinfo {volume} {12}},\
  \bibinfo {pages} {233} (\bibinfo {year} {2021})}\BibitemShut {NoStop}%
\bibitem [{\citenamefont {Harty}\ \emph {et~al.}(2014)\citenamefont {Harty},
  \citenamefont {Allcock}, \citenamefont {Ballance}, \citenamefont {Guidoni},
  \citenamefont {Janacek}, \citenamefont {Linke}, \citenamefont {Stacey},\ and\
  \citenamefont {Lucas}}]{PhysRevLett.113.220501}%
  \BibitemOpen
  \bibfield  {author} {\bibinfo {author} {\bibfnamefont {T.~P.}\ \bibnamefont
  {Harty}}, \bibinfo {author} {\bibfnamefont {D.~T.~C.}\ \bibnamefont
  {Allcock}}, \bibinfo {author} {\bibfnamefont {C.~J.}\ \bibnamefont
  {Ballance}}, \bibinfo {author} {\bibfnamefont {L.}~\bibnamefont {Guidoni}},
  \bibinfo {author} {\bibfnamefont {H.~A.}\ \bibnamefont {Janacek}}, \bibinfo
  {author} {\bibfnamefont {N.~M.}\ \bibnamefont {Linke}}, \bibinfo {author}
  {\bibfnamefont {D.~N.}\ \bibnamefont {Stacey}}, \ and\ \bibinfo {author}
  {\bibfnamefont {D.~M.}\ \bibnamefont {Lucas}},\ }\bibfield  {title} {\enquote
  {\bibinfo {title} {High-fidelity preparation, gates, memory, and readout of a
  trapped-ion quantum bit},}\ }\href {\doibase 10.1103/PhysRevLett.113.220501}
  {\bibfield  {journal} {\bibinfo  {journal} {Phys. Rev. Lett.}\ }\textbf
  {\bibinfo {volume} {113}},\ \bibinfo {pages} {220501} (\bibinfo {year}
  {2014})}\BibitemShut {NoStop}%
\bibitem [{\citenamefont {Ballance}\ \emph {et~al.}(2016)\citenamefont
  {Ballance}, \citenamefont {Harty}, \citenamefont {Linke}, \citenamefont
  {Sepiol},\ and\ \citenamefont {Lucas}}]{PhysRevLett.117.060504}%
  \BibitemOpen
  \bibfield  {author} {\bibinfo {author} {\bibfnamefont {C.~J.}\ \bibnamefont
  {Ballance}}, \bibinfo {author} {\bibfnamefont {T.~P.}\ \bibnamefont {Harty}},
  \bibinfo {author} {\bibfnamefont {N.~M.}\ \bibnamefont {Linke}}, \bibinfo
  {author} {\bibfnamefont {M.~A.}\ \bibnamefont {Sepiol}}, \ and\ \bibinfo
  {author} {\bibfnamefont {D.~M.}\ \bibnamefont {Lucas}},\ }\bibfield  {title}
  {\enquote {\bibinfo {title} {High-fidelity quantum logic gates using
  trapped-ion hyperfine qubits},}\ }\href {\doibase
  10.1103/PhysRevLett.117.060504} {\bibfield  {journal} {\bibinfo  {journal}
  {Phys. Rev. Lett.}\ }\textbf {\bibinfo {volume} {117}},\ \bibinfo {pages}
  {060504} (\bibinfo {year} {2016})}\BibitemShut {NoStop}%
\bibitem [{\citenamefont {Gaebler}\ \emph {et~al.}(2016)\citenamefont
  {Gaebler}, \citenamefont {Tan}, \citenamefont {Lin}, \citenamefont {Wan},
  \citenamefont {Bowler}, \citenamefont {Keith}, \citenamefont {Glancy},
  \citenamefont {Coakley}, \citenamefont {Knill}, \citenamefont {Leibfried},\
  and\ \citenamefont {Wineland}}]{PhysRevLett.117.060505}%
  \BibitemOpen
  \bibfield  {author} {\bibinfo {author} {\bibfnamefont {J.~P.}\ \bibnamefont
  {Gaebler}}, \bibinfo {author} {\bibfnamefont {T.~R.}\ \bibnamefont {Tan}},
  \bibinfo {author} {\bibfnamefont {Y.}~\bibnamefont {Lin}}, \bibinfo {author}
  {\bibfnamefont {Y.}~\bibnamefont {Wan}}, \bibinfo {author} {\bibfnamefont
  {R.}~\bibnamefont {Bowler}}, \bibinfo {author} {\bibfnamefont {A.~C.}\
  \bibnamefont {Keith}}, \bibinfo {author} {\bibfnamefont {S.}~\bibnamefont
  {Glancy}}, \bibinfo {author} {\bibfnamefont {K.}~\bibnamefont {Coakley}},
  \bibinfo {author} {\bibfnamefont {E.}~\bibnamefont {Knill}}, \bibinfo
  {author} {\bibfnamefont {D.}~\bibnamefont {Leibfried}}, \ and\ \bibinfo
  {author} {\bibfnamefont {D.~J.}\ \bibnamefont {Wineland}},\ }\bibfield
  {title} {\enquote {\bibinfo {title} {High-fidelity universal gate set for
  ${^{9}\mathrm{Be}}^{+}$ ion qubits},}\ }\href {\doibase
  10.1103/PhysRevLett.117.060505} {\bibfield  {journal} {\bibinfo  {journal}
  {Phys. Rev. Lett.}\ }\textbf {\bibinfo {volume} {117}},\ \bibinfo {pages}
  {060505} (\bibinfo {year} {2016})}\BibitemShut {NoStop}%
\bibitem [{\citenamefont {Clark}\ \emph {et~al.}(2021)\citenamefont {Clark},
  \citenamefont {Tinkey}, \citenamefont {Sawyer}, \citenamefont {Meier},
  \citenamefont {Burkhardt}, \citenamefont {Seck}, \citenamefont {Shappert},
  \citenamefont {Guise}, \citenamefont {Volin}, \citenamefont {Fallek},
  \citenamefont {Hayden}, \citenamefont {Rellergert},\ and\ \citenamefont
  {Brown}}]{PhysRevLett.127.130505}%
  \BibitemOpen
  \bibfield  {author} {\bibinfo {author} {\bibfnamefont {Craig~R.}\
  \bibnamefont {Clark}}, \bibinfo {author} {\bibfnamefont {Holly~N.}\
  \bibnamefont {Tinkey}}, \bibinfo {author} {\bibfnamefont {Brian~C.}\
  \bibnamefont {Sawyer}}, \bibinfo {author} {\bibfnamefont {Adam~M.}\
  \bibnamefont {Meier}}, \bibinfo {author} {\bibfnamefont {Karl~A.}\
  \bibnamefont {Burkhardt}}, \bibinfo {author} {\bibfnamefont {Christopher~M.}\
  \bibnamefont {Seck}}, \bibinfo {author} {\bibfnamefont {Christopher~M.}\
  \bibnamefont {Shappert}}, \bibinfo {author} {\bibfnamefont {Nicholas~D.}\
  \bibnamefont {Guise}}, \bibinfo {author} {\bibfnamefont {Curtis~E.}\
  \bibnamefont {Volin}}, \bibinfo {author} {\bibfnamefont {Spencer~D.}\
  \bibnamefont {Fallek}}, \bibinfo {author} {\bibfnamefont {Harley~T.}\
  \bibnamefont {Hayden}}, \bibinfo {author} {\bibfnamefont {Wade~G.}\
  \bibnamefont {Rellergert}}, \ and\ \bibinfo {author} {\bibfnamefont
  {Kenton~R.}\ \bibnamefont {Brown}},\ }\bibfield  {title} {\enquote {\bibinfo
  {title} {High-fidelity bell-state preparation with $^{40}{\mathrm{ca}}^{+}$
  optical qubits},}\ }\href {\doibase 10.1103/PhysRevLett.127.130505}
  {\bibfield  {journal} {\bibinfo  {journal} {Phys. Rev. Lett.}\ }\textbf
  {\bibinfo {volume} {127}},\ \bibinfo {pages} {130505} (\bibinfo {year}
  {2021})}\BibitemShut {NoStop}%
\bibitem [{\citenamefont {Blinov}\ \emph {et~al.}(2004)\citenamefont {Blinov},
  \citenamefont {Moehring}, \citenamefont {Duan},\ and\ \citenamefont
  {Monroe}}]{Blinov2004}%
  \BibitemOpen
  \bibfield  {author} {\bibinfo {author} {\bibfnamefont {B.~B.}\ \bibnamefont
  {Blinov}}, \bibinfo {author} {\bibfnamefont {D.~L.}\ \bibnamefont
  {Moehring}}, \bibinfo {author} {\bibfnamefont {L.-.~M.}\ \bibnamefont
  {Duan}}, \ and\ \bibinfo {author} {\bibfnamefont {C.}~\bibnamefont
  {Monroe}},\ }\bibfield  {title} {\enquote {\bibinfo {title} {Observation of
  entanglement between a single trapped atom and a single photon},}\ }\href
  {\doibase 10.1038/nature02377} {\bibfield  {journal} {\bibinfo  {journal}
  {Nature}\ }\textbf {\bibinfo {volume} {428}},\ \bibinfo {pages} {153--157}
  (\bibinfo {year} {2004})}\BibitemShut {NoStop}%
\bibitem [{\citenamefont {Hucul}\ \emph {et~al.}(2015)\citenamefont {Hucul},
  \citenamefont {Inlek}, \citenamefont {Vittorini}, \citenamefont {Crocker},
  \citenamefont {Debnath}, \citenamefont {Clark},\ and\ \citenamefont
  {Monroe}}]{Hucul2015}%
  \BibitemOpen
  \bibfield  {author} {\bibinfo {author} {\bibfnamefont {D.}~\bibnamefont
  {Hucul}}, \bibinfo {author} {\bibfnamefont {I.~V.}\ \bibnamefont {Inlek}},
  \bibinfo {author} {\bibfnamefont {G.}~\bibnamefont {Vittorini}}, \bibinfo
  {author} {\bibfnamefont {C.}~\bibnamefont {Crocker}}, \bibinfo {author}
  {\bibfnamefont {S.}~\bibnamefont {Debnath}}, \bibinfo {author} {\bibfnamefont
  {S.~M.}\ \bibnamefont {Clark}}, \ and\ \bibinfo {author} {\bibfnamefont
  {C.}~\bibnamefont {Monroe}},\ }\bibfield  {title} {\enquote {\bibinfo {title}
  {Modular entanglement of atomic qubits using photons and phonons},}\ }\href
  {\doibase 10.1038/nphys3150} {\bibfield  {journal} {\bibinfo  {journal}
  {Nature Physics}\ }\textbf {\bibinfo {volume} {11}},\ \bibinfo {pages}
  {37--42} (\bibinfo {year} {2015})}\BibitemShut {NoStop}%
\bibitem [{\citenamefont {Stephenson}\ \emph {et~al.}(2020)\citenamefont
  {Stephenson}, \citenamefont {Nadlinger}, \citenamefont {Nichol},
  \citenamefont {An}, \citenamefont {Drmota}, \citenamefont {Ballance},
  \citenamefont {Thirumalai}, \citenamefont {Goodwin}, \citenamefont {Lucas},\
  and\ \citenamefont {Ballance}}]{PhysRevLett.124.110501}%
  \BibitemOpen
  \bibfield  {author} {\bibinfo {author} {\bibfnamefont {L.~J.}\ \bibnamefont
  {Stephenson}}, \bibinfo {author} {\bibfnamefont {D.~P.}\ \bibnamefont
  {Nadlinger}}, \bibinfo {author} {\bibfnamefont {B.~C.}\ \bibnamefont
  {Nichol}}, \bibinfo {author} {\bibfnamefont {S.}~\bibnamefont {An}}, \bibinfo
  {author} {\bibfnamefont {P.}~\bibnamefont {Drmota}}, \bibinfo {author}
  {\bibfnamefont {T.~G.}\ \bibnamefont {Ballance}}, \bibinfo {author}
  {\bibfnamefont {K.}~\bibnamefont {Thirumalai}}, \bibinfo {author}
  {\bibfnamefont {J.~F.}\ \bibnamefont {Goodwin}}, \bibinfo {author}
  {\bibfnamefont {D.~M.}\ \bibnamefont {Lucas}}, \ and\ \bibinfo {author}
  {\bibfnamefont {C.~J.}\ \bibnamefont {Ballance}},\ }\bibfield  {title}
  {\enquote {\bibinfo {title} {High-rate, high-fidelity entanglement of qubits
  across an elementary quantum network},}\ }\href {\doibase
  10.1103/PhysRevLett.124.110501} {\bibfield  {journal} {\bibinfo  {journal}
  {Phys. Rev. Lett.}\ }\textbf {\bibinfo {volume} {124}},\ \bibinfo {pages}
  {110501} (\bibinfo {year} {2020})}\BibitemShut {NoStop}%
\bibitem [{\citenamefont {Bock}\ \emph {et~al.}(2018)\citenamefont {Bock},
  \citenamefont {Eich}, \citenamefont {Kucera}, \citenamefont {Kreis},
  \citenamefont {Lenhard}, \citenamefont {Becher},\ and\ \citenamefont
  {Eschner}}]{Bock2018}%
  \BibitemOpen
  \bibfield  {author} {\bibinfo {author} {\bibfnamefont {Matthias}\
  \bibnamefont {Bock}}, \bibinfo {author} {\bibfnamefont {Pascal}\ \bibnamefont
  {Eich}}, \bibinfo {author} {\bibfnamefont {Stephan}\ \bibnamefont {Kucera}},
  \bibinfo {author} {\bibfnamefont {Matthias}\ \bibnamefont {Kreis}}, \bibinfo
  {author} {\bibfnamefont {Andreas}\ \bibnamefont {Lenhard}}, \bibinfo {author}
  {\bibfnamefont {Christoph}\ \bibnamefont {Becher}}, \ and\ \bibinfo {author}
  {\bibfnamefont {J{\"u}rgen}\ \bibnamefont {Eschner}},\ }\bibfield  {title}
  {\enquote {\bibinfo {title} {High-fidelity entanglement between a trapped ion
  and a telecom photon via quantum frequency conversion},}\ }\href {\doibase
  10.1038/s41467-018-04341-2} {\bibfield  {journal} {\bibinfo  {journal}
  {Nature Communications}\ }\textbf {\bibinfo {volume} {9}},\ \bibinfo {pages}
  {1998} (\bibinfo {year} {2018})}\BibitemShut {NoStop}%
\bibitem [{\citenamefont {Krutyanskiy}\ \emph {et~al.}(2023)\citenamefont
  {Krutyanskiy}, \citenamefont {Galli}, \citenamefont {Krcmarsky},
  \citenamefont {Baier}, \citenamefont {Fioretto}, \citenamefont {Pu},
  \citenamefont {Mazloom}, \citenamefont {Sekatski}, \citenamefont {Canteri},
  \citenamefont {Teller}, \citenamefont {Schupp}, \citenamefont {Bate},
  \citenamefont {Meraner}, \citenamefont {Sangouard}, \citenamefont {Lanyon},\
  and\ \citenamefont {Northup}}]{PhysRevLett.130.050803}%
  \BibitemOpen
  \bibfield  {author} {\bibinfo {author} {\bibfnamefont {V.}~\bibnamefont
  {Krutyanskiy}}, \bibinfo {author} {\bibfnamefont {M.}~\bibnamefont {Galli}},
  \bibinfo {author} {\bibfnamefont {V.}~\bibnamefont {Krcmarsky}}, \bibinfo
  {author} {\bibfnamefont {S.}~\bibnamefont {Baier}}, \bibinfo {author}
  {\bibfnamefont {D.~A.}\ \bibnamefont {Fioretto}}, \bibinfo {author}
  {\bibfnamefont {Y.}~\bibnamefont {Pu}}, \bibinfo {author} {\bibfnamefont
  {A.}~\bibnamefont {Mazloom}}, \bibinfo {author} {\bibfnamefont
  {P.}~\bibnamefont {Sekatski}}, \bibinfo {author} {\bibfnamefont
  {M.}~\bibnamefont {Canteri}}, \bibinfo {author} {\bibfnamefont
  {M.}~\bibnamefont {Teller}}, \bibinfo {author} {\bibfnamefont
  {J.}~\bibnamefont {Schupp}}, \bibinfo {author} {\bibfnamefont
  {J.}~\bibnamefont {Bate}}, \bibinfo {author} {\bibfnamefont {M.}~\bibnamefont
  {Meraner}}, \bibinfo {author} {\bibfnamefont {N.}~\bibnamefont {Sangouard}},
  \bibinfo {author} {\bibfnamefont {B.~P.}\ \bibnamefont {Lanyon}}, \ and\
  \bibinfo {author} {\bibfnamefont {T.~E.}\ \bibnamefont {Northup}},\
  }\bibfield  {title} {\enquote {\bibinfo {title} {Entanglement of trapped-ion
  qubits separated by 230 meters},}\ }\href {\doibase
  10.1103/PhysRevLett.130.050803} {\bibfield  {journal} {\bibinfo  {journal}
  {Phys. Rev. Lett.}\ }\textbf {\bibinfo {volume} {130}},\ \bibinfo {pages}
  {050803} (\bibinfo {year} {2023})}\BibitemShut {NoStop}%
\bibitem [{\citenamefont {Feng}\ \emph {et~al.}(2024)\citenamefont {Feng},
  \citenamefont {Huang}, \citenamefont {Wu}, \citenamefont {Guo}, \citenamefont
  {Ma}, \citenamefont {Yang}, \citenamefont {Zhang}, \citenamefont {Wang},
  \citenamefont {Huang}, \citenamefont {Zhang}, \citenamefont {Yao},
  \citenamefont {Qi}, \citenamefont {Pu}, \citenamefont {Zhou},\ and\
  \citenamefont {Duan}}]{Feng2024}%
  \BibitemOpen
  \bibfield  {author} {\bibinfo {author} {\bibfnamefont {L.}~\bibnamefont
  {Feng}}, \bibinfo {author} {\bibfnamefont {Y.-Y.}\ \bibnamefont {Huang}},
  \bibinfo {author} {\bibfnamefont {Y.-K.}\ \bibnamefont {Wu}}, \bibinfo
  {author} {\bibfnamefont {W.-X.}\ \bibnamefont {Guo}}, \bibinfo {author}
  {\bibfnamefont {J.-Y.}\ \bibnamefont {Ma}}, \bibinfo {author} {\bibfnamefont
  {H.-X.}\ \bibnamefont {Yang}}, \bibinfo {author} {\bibfnamefont
  {L.}~\bibnamefont {Zhang}}, \bibinfo {author} {\bibfnamefont
  {Y.}~\bibnamefont {Wang}}, \bibinfo {author} {\bibfnamefont {C.-X.}\
  \bibnamefont {Huang}}, \bibinfo {author} {\bibfnamefont {C.}~\bibnamefont
  {Zhang}}, \bibinfo {author} {\bibfnamefont {L.}~\bibnamefont {Yao}}, \bibinfo
  {author} {\bibfnamefont {B.-X.}\ \bibnamefont {Qi}}, \bibinfo {author}
  {\bibfnamefont {Y.-F.}\ \bibnamefont {Pu}}, \bibinfo {author} {\bibfnamefont
  {Z.-C.}\ \bibnamefont {Zhou}}, \ and\ \bibinfo {author} {\bibfnamefont
  {L.-M.}\ \bibnamefont {Duan}},\ }\bibfield  {title} {\enquote {\bibinfo
  {title} {Realization of a crosstalk-avoided quantum network node using
  dual-type qubits of the same ion species},}\ }\href {\doibase
  10.1038/s41467-023-44220-z} {\bibfield  {journal} {\bibinfo  {journal}
  {Nature Communications}\ }\textbf {\bibinfo {volume} {15}},\ \bibinfo {pages}
  {204} (\bibinfo {year} {2024})}\BibitemShut {NoStop}%
\bibitem [{\citenamefont {Inlek}\ \emph {et~al.}(2017)\citenamefont {Inlek},
  \citenamefont {Crocker}, \citenamefont {Lichtman}, \citenamefont {Sosnova},\
  and\ \citenamefont {Monroe}}]{PhysRevLett.118.250502}%
  \BibitemOpen
  \bibfield  {author} {\bibinfo {author} {\bibfnamefont {I.~V.}\ \bibnamefont
  {Inlek}}, \bibinfo {author} {\bibfnamefont {C.}~\bibnamefont {Crocker}},
  \bibinfo {author} {\bibfnamefont {M.}~\bibnamefont {Lichtman}}, \bibinfo
  {author} {\bibfnamefont {K.}~\bibnamefont {Sosnova}}, \ and\ \bibinfo
  {author} {\bibfnamefont {C.}~\bibnamefont {Monroe}},\ }\bibfield  {title}
  {\enquote {\bibinfo {title} {Multispecies trapped-ion node for quantum
  networking},}\ }\href {\doibase 10.1103/PhysRevLett.118.250502} {\bibfield
  {journal} {\bibinfo  {journal} {Phys. Rev. Lett.}\ }\textbf {\bibinfo
  {volume} {118}},\ \bibinfo {pages} {250502} (\bibinfo {year}
  {2017})}\BibitemShut {NoStop}%
\bibitem [{\citenamefont {Drmota}\ \emph {et~al.}(2023)\citenamefont {Drmota},
  \citenamefont {Main}, \citenamefont {Nadlinger}, \citenamefont {Nichol},
  \citenamefont {Weber}, \citenamefont {Ainley}, \citenamefont {Agrawal},
  \citenamefont {Srinivas}, \citenamefont {Araneda}, \citenamefont {Ballance},\
  and\ \citenamefont {Lucas}}]{PhysRevLett.130.090803}%
  \BibitemOpen
  \bibfield  {author} {\bibinfo {author} {\bibfnamefont {P.}~\bibnamefont
  {Drmota}}, \bibinfo {author} {\bibfnamefont {D.}~\bibnamefont {Main}},
  \bibinfo {author} {\bibfnamefont {D.~P.}\ \bibnamefont {Nadlinger}}, \bibinfo
  {author} {\bibfnamefont {B.~C.}\ \bibnamefont {Nichol}}, \bibinfo {author}
  {\bibfnamefont {M.~A.}\ \bibnamefont {Weber}}, \bibinfo {author}
  {\bibfnamefont {E.~M.}\ \bibnamefont {Ainley}}, \bibinfo {author}
  {\bibfnamefont {A.}~\bibnamefont {Agrawal}}, \bibinfo {author} {\bibfnamefont
  {R.}~\bibnamefont {Srinivas}}, \bibinfo {author} {\bibfnamefont
  {G.}~\bibnamefont {Araneda}}, \bibinfo {author} {\bibfnamefont {C.~J.}\
  \bibnamefont {Ballance}}, \ and\ \bibinfo {author} {\bibfnamefont {D.~M.}\
  \bibnamefont {Lucas}},\ }\bibfield  {title} {\enquote {\bibinfo {title}
  {Robust quantum memory in a trapped-ion quantum network node},}\ }\href
  {\doibase 10.1103/PhysRevLett.130.090803} {\bibfield  {journal} {\bibinfo
  {journal} {Phys. Rev. Lett.}\ }\textbf {\bibinfo {volume} {130}},\ \bibinfo
  {pages} {090803} (\bibinfo {year} {2023})}\BibitemShut {NoStop}%
\bibitem [{\citenamefont {Main}\ \emph {et~al.}(2025)\citenamefont {Main},
  \citenamefont {Drmota}, \citenamefont {Nadlinger}, \citenamefont {Ainley},
  \citenamefont {Agrawal}, \citenamefont {Nichol}, \citenamefont {Srinivas},
  \citenamefont {Araneda},\ and\ \citenamefont {Lucas}}]{Main2025}%
  \BibitemOpen
  \bibfield  {author} {\bibinfo {author} {\bibfnamefont {D.}~\bibnamefont
  {Main}}, \bibinfo {author} {\bibfnamefont {P.}~\bibnamefont {Drmota}},
  \bibinfo {author} {\bibfnamefont {D.~P.}\ \bibnamefont {Nadlinger}}, \bibinfo
  {author} {\bibfnamefont {E.~M.}\ \bibnamefont {Ainley}}, \bibinfo {author}
  {\bibfnamefont {A.}~\bibnamefont {Agrawal}}, \bibinfo {author} {\bibfnamefont
  {B.~C.}\ \bibnamefont {Nichol}}, \bibinfo {author} {\bibfnamefont
  {R.}~\bibnamefont {Srinivas}}, \bibinfo {author} {\bibfnamefont
  {G.}~\bibnamefont {Araneda}}, \ and\ \bibinfo {author} {\bibfnamefont
  {D.~M.}\ \bibnamefont {Lucas}},\ }\bibfield  {title} {\enquote {\bibinfo
  {title} {Distributed quantum computing across an optical network link},}\
  }\href {\doibase 10.1038/s41586-024-08404-x} {\bibfield  {journal} {\bibinfo
  {journal} {Nature}\ }\textbf {\bibinfo {volume} {638}},\ \bibinfo {pages}
  {383--388} (\bibinfo {year} {2025})}\BibitemShut {NoStop}%
\bibitem [{\citenamefont {Sosnova}\ \emph {et~al.}(2021)\citenamefont
  {Sosnova}, \citenamefont {Carter},\ and\ \citenamefont
  {Monroe}}]{PhysRevA.103.012610}%
  \BibitemOpen
  \bibfield  {author} {\bibinfo {author} {\bibfnamefont {K.}~\bibnamefont
  {Sosnova}}, \bibinfo {author} {\bibfnamefont {A.}~\bibnamefont {Carter}}, \
  and\ \bibinfo {author} {\bibfnamefont {C.}~\bibnamefont {Monroe}},\
  }\bibfield  {title} {\enquote {\bibinfo {title} {Character of motional modes
  for entanglement and sympathetic cooling of mixed-species trapped-ion
  chains},}\ }\href {\doibase 10.1103/PhysRevA.103.012610} {\bibfield
  {journal} {\bibinfo  {journal} {Phys. Rev. A}\ }\textbf {\bibinfo {volume}
  {103}},\ \bibinfo {pages} {012610} (\bibinfo {year} {2021})}\BibitemShut
  {NoStop}%
\bibitem [{\citenamefont {Kielpinski}\ \emph {et~al.}(2002)\citenamefont
  {Kielpinski}, \citenamefont {Monroe},\ and\ \citenamefont
  {Wineland}}]{Kielpinski2002}%
  \BibitemOpen
  \bibfield  {author} {\bibinfo {author} {\bibfnamefont {D.}~\bibnamefont
  {Kielpinski}}, \bibinfo {author} {\bibfnamefont {C.}~\bibnamefont {Monroe}},
  \ and\ \bibinfo {author} {\bibfnamefont {D.~J.}\ \bibnamefont {Wineland}},\
  }\bibfield  {title} {\enquote {\bibinfo {title} {Architecture for a
  large-scale ion-trap quantum computer},}\ }\href {\doibase
  10.1038/nature00784} {\bibfield  {journal} {\bibinfo  {journal} {Nature}\
  }\textbf {\bibinfo {volume} {417}},\ \bibinfo {pages} {709--711} (\bibinfo
  {year} {2002})}\BibitemShut {NoStop}%
\bibitem [{\citenamefont {You}\ \emph {et~al.}(2024)\citenamefont {You},
  \citenamefont {Wu}, \citenamefont {Miron}, \citenamefont {Ke}, \citenamefont
  {Monga}, \citenamefont {Saglamyurek},\ and\ \citenamefont
  {Haeffner}}]{you2024temporally}%
  \BibitemOpen
  \bibfield  {author} {\bibinfo {author} {\bibfnamefont {Bingran}\ \bibnamefont
  {You}}, \bibinfo {author} {\bibfnamefont {Qiming}\ \bibnamefont {Wu}},
  \bibinfo {author} {\bibfnamefont {David}\ \bibnamefont {Miron}}, \bibinfo
  {author} {\bibfnamefont {Wenjun}\ \bibnamefont {Ke}}, \bibinfo {author}
  {\bibfnamefont {Inder}\ \bibnamefont {Monga}}, \bibinfo {author}
  {\bibfnamefont {Erhan}\ \bibnamefont {Saglamyurek}}, \ and\ \bibinfo {author}
  {\bibfnamefont {Hartmut}\ \bibnamefont {Haeffner}},\ }\bibfield  {title}
  {\enquote {\bibinfo {title} {Temporally multiplexed ion-photon quantum
  interface via fast ion-chain transport},}\ }\href
  {https://arxiv.org/abs/2405.10501} {\bibfield  {journal} {\bibinfo  {journal}
  {arXiv:2405.10501}\ } (\bibinfo {year} {2024})}\BibitemShut {NoStop}%
\bibitem [{\citenamefont {Cui}\ \emph {et~al.}(2025)\citenamefont {Cui},
  \citenamefont {Wang}, \citenamefont {Lai}, \citenamefont {Wang},
  \citenamefont {Shi}, \citenamefont {Liu}, \citenamefont {Sun}, \citenamefont
  {Tian}, \citenamefont {Liang}, \citenamefont {Qi}, \citenamefont {Huang},
  \citenamefont {Zhou}, \citenamefont {Wu}, \citenamefont {Xu}, \citenamefont
  {Duan},\ and\ \citenamefont {Pu}}]{cui2025metropolitan}%
  \BibitemOpen
  \bibfield  {author} {\bibinfo {author} {\bibfnamefont {Z.~B.}\ \bibnamefont
  {Cui}}, \bibinfo {author} {\bibfnamefont {Z.~Q.}\ \bibnamefont {Wang}},
  \bibinfo {author} {\bibfnamefont {P.~C.}\ \bibnamefont {Lai}}, \bibinfo
  {author} {\bibfnamefont {Y.}~\bibnamefont {Wang}}, \bibinfo {author}
  {\bibfnamefont {J.~X.}\ \bibnamefont {Shi}}, \bibinfo {author} {\bibfnamefont
  {P.~Y.}\ \bibnamefont {Liu}}, \bibinfo {author} {\bibfnamefont {Y.~D.}\
  \bibnamefont {Sun}}, \bibinfo {author} {\bibfnamefont {Z.~C.}\ \bibnamefont
  {Tian}}, \bibinfo {author} {\bibfnamefont {Y.~B.}\ \bibnamefont {Liang}},
  \bibinfo {author} {\bibfnamefont {B.~X.}\ \bibnamefont {Qi}}, \bibinfo
  {author} {\bibfnamefont {Y.~Y.}\ \bibnamefont {Huang}}, \bibinfo {author}
  {\bibfnamefont {Z.~C.}\ \bibnamefont {Zhou}}, \bibinfo {author}
  {\bibfnamefont {Y.~K.}\ \bibnamefont {Wu}}, \bibinfo {author} {\bibfnamefont
  {Y.}~\bibnamefont {Xu}}, \bibinfo {author} {\bibfnamefont {L.~M.}\
  \bibnamefont {Duan}}, \ and\ \bibinfo {author} {\bibfnamefont {Y.~F.}\
  \bibnamefont {Pu}},\ }\bibfield  {title} {\enquote {\bibinfo {title} {A
  metropolitan-scale trapped-ion quantum network node with hybrid multiplexing
  enhancements},}\ }\href {https://arxiv.org/abs/2503.13898} {\bibfield
  {journal} {\bibinfo  {journal} {arXiv:2503.13898}\ } (\bibinfo {year}
  {2025})}\BibitemShut {NoStop}%
\bibitem [{\citenamefont {Yang}\ \emph {et~al.}(2022)\citenamefont {Yang},
  \citenamefont {Ma}, \citenamefont {Wu}, \citenamefont {Wang}, \citenamefont
  {Cao}, \citenamefont {Guo}, \citenamefont {Huang}, \citenamefont {Feng},
  \citenamefont {Zhou},\ and\ \citenamefont {Duan}}]{yang2022realizing}%
  \BibitemOpen
  \bibfield  {author} {\bibinfo {author} {\bibfnamefont {H.-X.}\ \bibnamefont
  {Yang}}, \bibinfo {author} {\bibfnamefont {J.-Y.}\ \bibnamefont {Ma}},
  \bibinfo {author} {\bibfnamefont {Y.-K.}\ \bibnamefont {Wu}}, \bibinfo
  {author} {\bibfnamefont {Y.}~\bibnamefont {Wang}}, \bibinfo {author}
  {\bibfnamefont {M.-M.}\ \bibnamefont {Cao}}, \bibinfo {author} {\bibfnamefont
  {W.-X.}\ \bibnamefont {Guo}}, \bibinfo {author} {\bibfnamefont {Y.-Y.}\
  \bibnamefont {Huang}}, \bibinfo {author} {\bibfnamefont {L.}~\bibnamefont
  {Feng}}, \bibinfo {author} {\bibfnamefont {Z.-C.}\ \bibnamefont {Zhou}}, \
  and\ \bibinfo {author} {\bibfnamefont {L.-M.}\ \bibnamefont {Duan}},\
  }\bibfield  {title} {\enquote {\bibinfo {title} {Realizing coherently
  convertible dual-type qubits with the same ion species},}\ }\href {\doibase
  10.1038/s41567-022-01661-5} {\bibfield  {journal} {\bibinfo  {journal}
  {Nature Physics}\ }\textbf {\bibinfo {volume} {18}},\ \bibinfo {pages}
  {1058--1061} (\bibinfo {year} {2022})},\ \bibinfo {note}
  {arXiv:2106.14906}\BibitemShut {NoStop}%
\bibitem [{\citenamefont {Allcock}\ \emph {et~al.}(2021)\citenamefont
  {Allcock}, \citenamefont {Campbell}, \citenamefont {Chiaverini},
  \citenamefont {Chuang}, \citenamefont {Hudson}, \citenamefont {Moore},
  \citenamefont {Ransford}, \citenamefont {Roman}, \citenamefont {Sage},\ and\
  \citenamefont {Wineland}}]{10.1063/5.0069544}%
  \BibitemOpen
  \bibfield  {author} {\bibinfo {author} {\bibfnamefont {D.~T.~C.}\
  \bibnamefont {Allcock}}, \bibinfo {author} {\bibfnamefont {W.~C.}\
  \bibnamefont {Campbell}}, \bibinfo {author} {\bibfnamefont {J.}~\bibnamefont
  {Chiaverini}}, \bibinfo {author} {\bibfnamefont {I.~L.}\ \bibnamefont
  {Chuang}}, \bibinfo {author} {\bibfnamefont {E.~R.}\ \bibnamefont {Hudson}},
  \bibinfo {author} {\bibfnamefont {I.~D.}\ \bibnamefont {Moore}}, \bibinfo
  {author} {\bibfnamefont {A.}~\bibnamefont {Ransford}}, \bibinfo {author}
  {\bibfnamefont {C.}~\bibnamefont {Roman}}, \bibinfo {author} {\bibfnamefont
  {J.~M.}\ \bibnamefont {Sage}}, \ and\ \bibinfo {author} {\bibfnamefont
  {D.~J.}\ \bibnamefont {Wineland}},\ }\bibfield  {title} {\enquote {\bibinfo
  {title} {{omg blueprint for trapped ion quantum computing with metastable
  states}},}\ }\href {\doibase 10.1063/5.0069544} {\bibfield  {journal}
  {\bibinfo  {journal} {Applied Physics Letters}\ }\textbf {\bibinfo {volume}
  {119}},\ \bibinfo {pages} {214002} (\bibinfo {year} {2021})},\ \bibinfo
  {note} {arXiv:2109.01272}\BibitemShut {NoStop}%
\bibitem [{\citenamefont {Lai}\ \emph {et~al.}(2025)\citenamefont {Lai},
  \citenamefont {Wang}, \citenamefont {Shi}, \citenamefont {Cui}, \citenamefont
  {Wang}, \citenamefont {Zhang}, \citenamefont {Liu}, \citenamefont {Tian},
  \citenamefont {Sun}, \citenamefont {Chang}, \citenamefont {Qi}, \citenamefont
  {Huang}, \citenamefont {Zhou}, \citenamefont {Wu}, \citenamefont {Xu},
  \citenamefont {Pu},\ and\ \citenamefont {Duan}}]{PhysRevLett.134.070801}%
  \BibitemOpen
  \bibfield  {author} {\bibinfo {author} {\bibfnamefont {P.-C.}\ \bibnamefont
  {Lai}}, \bibinfo {author} {\bibfnamefont {Y.}~\bibnamefont {Wang}}, \bibinfo
  {author} {\bibfnamefont {J.-X.}\ \bibnamefont {Shi}}, \bibinfo {author}
  {\bibfnamefont {Z.-B.}\ \bibnamefont {Cui}}, \bibinfo {author} {\bibfnamefont
  {Z.-Q.}\ \bibnamefont {Wang}}, \bibinfo {author} {\bibfnamefont
  {S.}~\bibnamefont {Zhang}}, \bibinfo {author} {\bibfnamefont {P.-Y.}\
  \bibnamefont {Liu}}, \bibinfo {author} {\bibfnamefont {Z.-C.}\ \bibnamefont
  {Tian}}, \bibinfo {author} {\bibfnamefont {Y.-D.}\ \bibnamefont {Sun}},
  \bibinfo {author} {\bibfnamefont {X.-Y.}\ \bibnamefont {Chang}}, \bibinfo
  {author} {\bibfnamefont {B.-X.}\ \bibnamefont {Qi}}, \bibinfo {author}
  {\bibfnamefont {Y.-Y.}\ \bibnamefont {Huang}}, \bibinfo {author}
  {\bibfnamefont {Z.-C.}\ \bibnamefont {Zhou}}, \bibinfo {author}
  {\bibfnamefont {Y.-K.}\ \bibnamefont {Wu}}, \bibinfo {author} {\bibfnamefont
  {Y.}~\bibnamefont {Xu}}, \bibinfo {author} {\bibfnamefont {Y.-F.}\
  \bibnamefont {Pu}}, \ and\ \bibinfo {author} {\bibfnamefont {L.-M.}\
  \bibnamefont {Duan}},\ }\bibfield  {title} {\enquote {\bibinfo {title}
  {Realization of a crosstalk-free two-ion node for long-distance quantum
  networking},}\ }\href {\doibase 10.1103/PhysRevLett.134.070801} {\bibfield
  {journal} {\bibinfo  {journal} {Phys. Rev. Lett.}\ }\textbf {\bibinfo
  {volume} {134}},\ \bibinfo {pages} {070801} (\bibinfo {year}
  {2025})}\BibitemShut {NoStop}%
\bibitem [{\citenamefont {Guo}\ \emph {et~al.}(2024)\citenamefont {Guo},
  \citenamefont {Wu}, \citenamefont {Ye}, \citenamefont {Zhang}, \citenamefont
  {Lian}, \citenamefont {Yao}, \citenamefont {Wang}, \citenamefont {Yan},
  \citenamefont {Yi}, \citenamefont {Xu}, \citenamefont {Li}, \citenamefont
  {Hou}, \citenamefont {Xu}, \citenamefont {Guo}, \citenamefont {Zhang},
  \citenamefont {Qi}, \citenamefont {Zhou}, \citenamefont {He},\ and\
  \citenamefont {Duan}}]{guo2024siteresolved}%
  \BibitemOpen
  \bibfield  {author} {\bibinfo {author} {\bibfnamefont {S.-A.}\ \bibnamefont
  {Guo}}, \bibinfo {author} {\bibfnamefont {Y.-K.}\ \bibnamefont {Wu}},
  \bibinfo {author} {\bibfnamefont {J.}~\bibnamefont {Ye}}, \bibinfo {author}
  {\bibfnamefont {L.}~\bibnamefont {Zhang}}, \bibinfo {author} {\bibfnamefont
  {W.-Q.}\ \bibnamefont {Lian}}, \bibinfo {author} {\bibfnamefont
  {R.}~\bibnamefont {Yao}}, \bibinfo {author} {\bibfnamefont {Y.}~\bibnamefont
  {Wang}}, \bibinfo {author} {\bibfnamefont {R.-Y.}\ \bibnamefont {Yan}},
  \bibinfo {author} {\bibfnamefont {Y.-J.}\ \bibnamefont {Yi}}, \bibinfo
  {author} {\bibfnamefont {Y.-L.}\ \bibnamefont {Xu}}, \bibinfo {author}
  {\bibfnamefont {B.-W.}\ \bibnamefont {Li}}, \bibinfo {author} {\bibfnamefont
  {Y.-H.}\ \bibnamefont {Hou}}, \bibinfo {author} {\bibfnamefont {Y.-Z.}\
  \bibnamefont {Xu}}, \bibinfo {author} {\bibfnamefont {W.-X.}\ \bibnamefont
  {Guo}}, \bibinfo {author} {\bibfnamefont {C.}~\bibnamefont {Zhang}}, \bibinfo
  {author} {\bibfnamefont {B.-X.}\ \bibnamefont {Qi}}, \bibinfo {author}
  {\bibfnamefont {Z.-C.}\ \bibnamefont {Zhou}}, \bibinfo {author}
  {\bibfnamefont {L.}~\bibnamefont {He}}, \ and\ \bibinfo {author}
  {\bibfnamefont {L.-M.}\ \bibnamefont {Duan}},\ }\bibfield  {title} {\enquote
  {\bibinfo {title} {A site-resolved two-dimensional quantum simulator with
  hundreds of trapped ions},}\ }\href {\doibase 10.1038/s41586-024-07459-0}
  {\bibfield  {journal} {\bibinfo  {journal} {Nature}\ }\textbf {\bibinfo
  {volume} {630}},\ \bibinfo {pages} {613--618} (\bibinfo {year}
  {2024})}\BibitemShut {NoStop}%
\bibitem [{\citenamefont {Hou}\ \emph {et~al.}(2024)\citenamefont {Hou},
  \citenamefont {Yi}, \citenamefont {Wu}, \citenamefont {Chen}, \citenamefont
  {Zhang}, \citenamefont {Wang}, \citenamefont {Xu}, \citenamefont {Zhang},
  \citenamefont {Mei}, \citenamefont {Yang}, \citenamefont {Ma}, \citenamefont
  {Guo}, \citenamefont {Ye}, \citenamefont {Qi}, \citenamefont {Zhou},
  \citenamefont {Hou},\ and\ \citenamefont {Duan}}]{Hou2024}%
  \BibitemOpen
  \bibfield  {author} {\bibinfo {author} {\bibfnamefont {Y.-H.}\ \bibnamefont
  {Hou}}, \bibinfo {author} {\bibfnamefont {Y.-J.}\ \bibnamefont {Yi}},
  \bibinfo {author} {\bibfnamefont {Y.-K.}\ \bibnamefont {Wu}}, \bibinfo
  {author} {\bibfnamefont {Y.-Y.}\ \bibnamefont {Chen}}, \bibinfo {author}
  {\bibfnamefont {L.}~\bibnamefont {Zhang}}, \bibinfo {author} {\bibfnamefont
  {Y.}~\bibnamefont {Wang}}, \bibinfo {author} {\bibfnamefont {Y.-L.}\
  \bibnamefont {Xu}}, \bibinfo {author} {\bibfnamefont {C.}~\bibnamefont
  {Zhang}}, \bibinfo {author} {\bibfnamefont {Q.-X.}\ \bibnamefont {Mei}},
  \bibinfo {author} {\bibfnamefont {H.-X.}\ \bibnamefont {Yang}}, \bibinfo
  {author} {\bibfnamefont {J.-Y.}\ \bibnamefont {Ma}}, \bibinfo {author}
  {\bibfnamefont {S.-A.}\ \bibnamefont {Guo}}, \bibinfo {author} {\bibfnamefont
  {J.}~\bibnamefont {Ye}}, \bibinfo {author} {\bibfnamefont {B.-X.}\
  \bibnamefont {Qi}}, \bibinfo {author} {\bibfnamefont {Z.-C.}\ \bibnamefont
  {Zhou}}, \bibinfo {author} {\bibfnamefont {P.-Y.}\ \bibnamefont {Hou}}, \
  and\ \bibinfo {author} {\bibfnamefont {L.-M.}\ \bibnamefont {Duan}},\
  }\bibfield  {title} {\enquote {\bibinfo {title} {Individually addressed
  entangling gates in a two-dimensional ion crystal},}\ }\href {\doibase
  10.1038/s41467-024-53405-z} {\bibfield  {journal} {\bibinfo  {journal}
  {Nature Communications}\ }\textbf {\bibinfo {volume} {15}},\ \bibinfo {pages}
  {9710} (\bibinfo {year} {2024})}\BibitemShut {NoStop}%
\bibitem [{\citenamefont {B\ifmmode \u{a}\else \u{a}\fi{}z\ifmmode~\u{a}\else
  \u{a}\fi{}van}\ \emph {et~al.}(2023)\citenamefont {B\ifmmode \u{a}\else
  \u{a}\fi{}z\ifmmode~\u{a}\else \u{a}\fi{}van}, \citenamefont {Saner},
  \citenamefont {Minder}, \citenamefont {Hughes}, \citenamefont {Sutherland},
  \citenamefont {Lucas}, \citenamefont {Srinivas},\ and\ \citenamefont
  {Ballance}}]{PhysRevA.107.022617}%
  \BibitemOpen
  \bibfield  {author} {\bibinfo {author} {\bibfnamefont {O.}~\bibnamefont
  {B\ifmmode \u{a}\else \u{a}\fi{}z\ifmmode~\u{a}\else \u{a}\fi{}van}},
  \bibinfo {author} {\bibfnamefont {S.}~\bibnamefont {Saner}}, \bibinfo
  {author} {\bibfnamefont {M.}~\bibnamefont {Minder}}, \bibinfo {author}
  {\bibfnamefont {A.~C.}\ \bibnamefont {Hughes}}, \bibinfo {author}
  {\bibfnamefont {R.~T.}\ \bibnamefont {Sutherland}}, \bibinfo {author}
  {\bibfnamefont {D.~M.}\ \bibnamefont {Lucas}}, \bibinfo {author}
  {\bibfnamefont {R.}~\bibnamefont {Srinivas}}, \ and\ \bibinfo {author}
  {\bibfnamefont {C.~J.}\ \bibnamefont {Ballance}},\ }\bibfield  {title}
  {\enquote {\bibinfo {title} {Synthesizing a ${\stackrel{\ifmmode \hat{}\else
  \^{}\fi{}}{\ensuremath{\sigma}}}_{z}$ spin-dependent force for optical,
  metastable, and ground-state trapped-ion qubits},}\ }\href {\doibase
  10.1103/PhysRevA.107.022617} {\bibfield  {journal} {\bibinfo  {journal}
  {Phys. Rev. A}\ }\textbf {\bibinfo {volume} {107}},\ \bibinfo {pages}
  {022617} (\bibinfo {year} {2023})}\BibitemShut {NoStop}%
\bibitem [{\citenamefont {Roos}(2008)}]{Roos_2008}%
  \BibitemOpen
  \bibfield  {author} {\bibinfo {author} {\bibfnamefont {Christian~F}\
  \bibnamefont {Roos}},\ }\bibfield  {title} {\enquote {\bibinfo {title} {Ion
  trap quantum gates with amplitude-modulated laser beams},}\ }\href {\doibase
  10.1088/1367-2630/10/1/013002} {\bibfield  {journal} {\bibinfo  {journal}
  {New Journal of Physics}\ }\textbf {\bibinfo {volume} {10}},\ \bibinfo
  {pages} {013002} (\bibinfo {year} {2008})}\BibitemShut {NoStop}%
\bibitem [{\citenamefont {Roman}\ \emph {et~al.}(2020)\citenamefont {Roman},
  \citenamefont {Ransford}, \citenamefont {Ip},\ and\ \citenamefont
  {Campbell}}]{Roman2020}%
  \BibitemOpen
  \bibfield  {author} {\bibinfo {author} {\bibfnamefont {Conrad}\ \bibnamefont
  {Roman}}, \bibinfo {author} {\bibfnamefont {Anthony}\ \bibnamefont
  {Ransford}}, \bibinfo {author} {\bibfnamefont {Michael}\ \bibnamefont {Ip}},
  \ and\ \bibinfo {author} {\bibfnamefont {Wesley~C}\ \bibnamefont
  {Campbell}},\ }\bibfield  {title} {\enquote {\bibinfo {title} {Coherent
  control for qubit state readout},}\ }\href {\doibase
  10.1088/1367-2630/ab9982} {\bibfield  {journal} {\bibinfo  {journal} {New
  Journal of Physics}\ }\textbf {\bibinfo {volume} {22}},\ \bibinfo {pages}
  {073038} (\bibinfo {year} {2020})}\BibitemShut {NoStop}%
\bibitem [{\citenamefont {Edmunds}\ \emph {et~al.}(2021)\citenamefont
  {Edmunds}, \citenamefont {Tan}, \citenamefont {Milne}, \citenamefont {Singh},
  \citenamefont {Biercuk},\ and\ \citenamefont {Hempel}}]{edmunds2020scalable}%
  \BibitemOpen
  \bibfield  {author} {\bibinfo {author} {\bibfnamefont {C.~L.}\ \bibnamefont
  {Edmunds}}, \bibinfo {author} {\bibfnamefont {T.~R.}\ \bibnamefont {Tan}},
  \bibinfo {author} {\bibfnamefont {A.~R.}\ \bibnamefont {Milne}}, \bibinfo
  {author} {\bibfnamefont {A.}~\bibnamefont {Singh}}, \bibinfo {author}
  {\bibfnamefont {M.~J.}\ \bibnamefont {Biercuk}}, \ and\ \bibinfo {author}
  {\bibfnamefont {C.}~\bibnamefont {Hempel}},\ }\bibfield  {title} {\enquote
  {\bibinfo {title} {Scalable hyperfine qubit state detection via electron
  shelving in the ${}^{2}{D}_{5/2}$ and ${}^{2}{F}_{7/2}$ manifolds in
  ${}^{171}{\mathrm{yb}}^{+}$},}\ }\href {\doibase 10.1103/PhysRevA.104.012606}
  {\bibfield  {journal} {\bibinfo  {journal} {Phys. Rev. A}\ }\textbf {\bibinfo
  {volume} {104}},\ \bibinfo {pages} {012606} (\bibinfo {year}
  {2021})}\BibitemShut {NoStop}%
\bibitem [{\citenamefont {Wootters}\ and\ \citenamefont
  {Fields}(1989)}]{WOOTTERS1989363}%
  \BibitemOpen
  \bibfield  {author} {\bibinfo {author} {\bibfnamefont {William~K}\
  \bibnamefont {Wootters}}\ and\ \bibinfo {author} {\bibfnamefont {Brian~D}\
  \bibnamefont {Fields}},\ }\bibfield  {title} {\enquote {\bibinfo {title}
  {Optimal state-determination by mutually unbiased measurements},}\ }\href
  {\doibase https://doi.org/10.1016/0003-4916(89)90322-9} {\bibfield  {journal}
  {\bibinfo  {journal} {Annals of Physics}\ }\textbf {\bibinfo {volume}
  {191}},\ \bibinfo {pages} {363--381} (\bibinfo {year} {1989})}\BibitemShut
  {NoStop}%
\bibitem [{\citenamefont {White}\ \emph {et~al.}(2007)\citenamefont {White},
  \citenamefont {Gilchrist}, \citenamefont {Pryde}, \citenamefont {O'Brien},
  \citenamefont {Bremner},\ and\ \citenamefont {Langford}}]{MeasureTwoQubits}%
  \BibitemOpen
  \bibfield  {author} {\bibinfo {author} {\bibfnamefont {Andrew~G.}\
  \bibnamefont {White}}, \bibinfo {author} {\bibfnamefont {Alexei}\
  \bibnamefont {Gilchrist}}, \bibinfo {author} {\bibfnamefont {Geoffrey~J.}\
  \bibnamefont {Pryde}}, \bibinfo {author} {\bibfnamefont {Jeremy~L.}\
  \bibnamefont {O'Brien}}, \bibinfo {author} {\bibfnamefont {Michael~J.}\
  \bibnamefont {Bremner}}, \ and\ \bibinfo {author} {\bibfnamefont {Nathan~K.}\
  \bibnamefont {Langford}},\ }\bibfield  {title} {\enquote {\bibinfo {title}
  {Measuring two-qubit gates},}\ }\href {\doibase 10.1364/JOSAB.24.000172}
  {\bibfield  {journal} {\bibinfo  {journal} {J. Opt. Soc. Am. B}\ }\textbf
  {\bibinfo {volume} {24}},\ \bibinfo {pages} {172--183} (\bibinfo {year}
  {2007})}\BibitemShut {NoStop}%
\bibitem [{\citenamefont {Gilchrist}\ \emph {et~al.}(2005)\citenamefont
  {Gilchrist}, \citenamefont {Langford},\ and\ \citenamefont
  {Nielsen}}]{PhysRevA.71.062310}%
  \BibitemOpen
  \bibfield  {author} {\bibinfo {author} {\bibfnamefont {Alexei}\ \bibnamefont
  {Gilchrist}}, \bibinfo {author} {\bibfnamefont {Nathan~K.}\ \bibnamefont
  {Langford}}, \ and\ \bibinfo {author} {\bibfnamefont {Michael~A.}\
  \bibnamefont {Nielsen}},\ }\bibfield  {title} {\enquote {\bibinfo {title}
  {Distance measures to compare real and ideal quantum processes},}\ }\href
  {\doibase 10.1103/PhysRevA.71.062310} {\bibfield  {journal} {\bibinfo
  {journal} {Phys. Rev. A}\ }\textbf {\bibinfo {volume} {71}},\ \bibinfo
  {pages} {062310} (\bibinfo {year} {2005})}\BibitemShut {NoStop}%
\bibitem [{\citenamefont {G\"uhne}\ \emph {et~al.}(2007)\citenamefont
  {G\"uhne}, \citenamefont {Lu}, \citenamefont {Gao},\ and\ \citenamefont
  {Pan}}]{PhysRevA.76.030305}%
  \BibitemOpen
  \bibfield  {author} {\bibinfo {author} {\bibfnamefont {Otfried}\ \bibnamefont
  {G\"uhne}}, \bibinfo {author} {\bibfnamefont {Chao-Yang}\ \bibnamefont {Lu}},
  \bibinfo {author} {\bibfnamefont {Wei-Bo}\ \bibnamefont {Gao}}, \ and\
  \bibinfo {author} {\bibfnamefont {Jian-Wei}\ \bibnamefont {Pan}},\ }\bibfield
   {title} {\enquote {\bibinfo {title} {Toolbox for entanglement detection and
  fidelity estimation},}\ }\href {\doibase 10.1103/PhysRevA.76.030305}
  {\bibfield  {journal} {\bibinfo  {journal} {Phys. Rev. A}\ }\textbf {\bibinfo
  {volume} {76}},\ \bibinfo {pages} {030305} (\bibinfo {year}
  {2007})}\BibitemShut {NoStop}%
\end{thebibliography}
%

\end{document}


\title{Supplementary Information for \\
``Realization of a functioning dual-type trapped-ion quantum network node''}

\author{Y.-Y. Huang}
\thanks{These authors contribute equally to this work}%
\affiliation{Center for Quantum Information, Institute for Interdisciplinary Information Sciences, Tsinghua University, Beijing 100084, PR China}

\author{L. Feng}
\thanks{These authors contribute equally to this work}%
\affiliation{Center for Quantum Information, Institute for Interdisciplinary Information Sciences, Tsinghua University, Beijing 100084, PR China}

\author{Y.-K. Wu}
\thanks{These authors contribute equally to this work}%
\affiliation{Center for Quantum Information, Institute for Interdisciplinary Information Sciences, Tsinghua University, Beijing 100084, PR China}
\affiliation{Hefei National Laboratory, Hefei 230088, PR China}

\author{Y.-L. Xu}
\affiliation{Center for Quantum Information, Institute for Interdisciplinary Information Sciences, Tsinghua University, Beijing 100084, PR China}

\author{L. Zhang}
\affiliation{Center for Quantum Information, Institute for Interdisciplinary Information Sciences, Tsinghua University, Beijing 100084, PR China}

\author{Z.-B. Cui}
\affiliation{Center for Quantum Information, Institute for Interdisciplinary Information Sciences, Tsinghua University, Beijing 100084, PR China}

\author{C.-X. Huang}
\affiliation{Center for Quantum Information, Institute for Interdisciplinary Information Sciences, Tsinghua University, Beijing 100084, PR China}

\author{C. Zhang}
\affiliation{HYQ Co., Ltd., Beijing 100176, P. R. China}

\author{S.-A. Guo}
\affiliation{Center for Quantum Information, Institute for Interdisciplinary Information Sciences, Tsinghua University, Beijing 100084, PR China}

\author{Q.-X. Mei}
\affiliation{HYQ Co., Ltd., Beijing 100176, P. R. China}

\author{B.-X. Qi}
\affiliation{Center for Quantum Information, Institute for Interdisciplinary Information Sciences, Tsinghua University, Beijing 100084, PR China}

\author{Y. Xu}
\affiliation{Center for Quantum Information, Institute for Interdisciplinary Information Sciences, Tsinghua University, Beijing 100084, PR China}
\affiliation{Hefei National Laboratory, Hefei 230088, PR China}

\author{Y.-F. Pu}
\affiliation{Center for Quantum Information, Institute for Interdisciplinary Information Sciences, Tsinghua University, Beijing 100084, PR China}
\affiliation{Hefei National Laboratory, Hefei 230088, PR China}

\author{Z.-C. Zhou}
\affiliation{Center for Quantum Information, Institute for Interdisciplinary Information Sciences, Tsinghua University, Beijing 100084, PR China}
\affiliation{Hefei National Laboratory, Hefei 230088, PR China}

\author{L.-M. Duan}
\email{lmduan@tsinghua.edu.cn}
\affiliation{Center for Quantum Information, Institute for Interdisciplinary Information Sciences, Tsinghua University, Beijing 100084, PR China}
\affiliation{Hefei National Laboratory, Hefei 230088, PR China}

\maketitle

\makeatletter
\renewcommand{\thefigure}{S\arabic{figure}}
\renewcommand{\thetable}{S\arabic{table}}
\renewcommand{\theequation}{S\arabic{equation}}
\makeatother

\section{Motional heating induced by ion-photon entanglement generation}
\begin{figure}[htbp]
\centering
\includegraphics[width=2.7in]{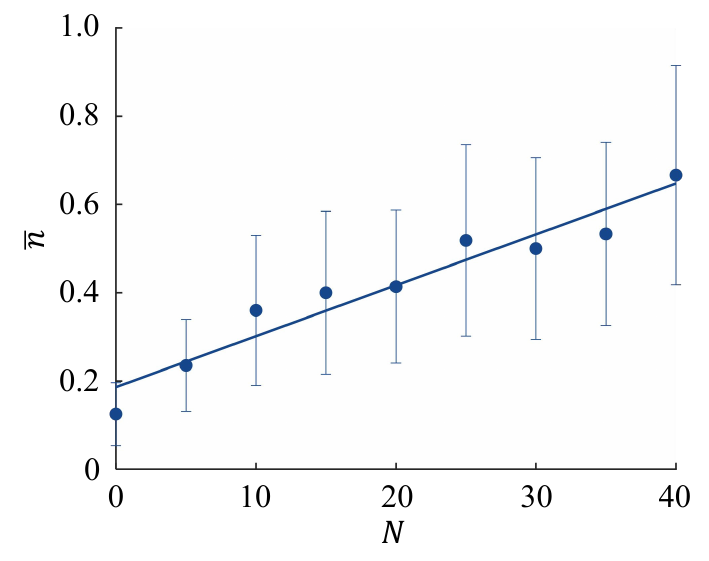}
\caption{\textbf{Motional heating induced by the ion-photon entanglement generation.} After Doppler cooling and EIT cooling, we repeat the entanglement generation attempt for $N$ rounds and then measure the average phonon number $\bar{n}$.}
\label{fig-S1}
\end{figure}

During the ion-photon entanglement generation stage, the ions are subjected to the motional heating due to the random scattering of the photons and other environmental noise. Although the heating effect for each generation attempt can be small, it will accumulate between multiple attempts and harm the later manipulation of the ionic qubits. In principle, we can perform Doppler cooling and EIT cooling before each entanglement generation attempt to avoid this problem. However, as we describe in the main text, the duration of the cooling stage is much longer than that of a single entanglement generation attempt. Therefore we choose to repeat up to $N=10$ attempts before we move back to the cooling stage, as shown in Fig.1\textbf{c} of the main text.
In Fig.~\ref{fig-S1} we measure the average phonon number $\bar{n}$ of the memory ion from the excitation probability ratio of the red and blue motional sidebands \cite{RevModPhys.75.281} under a strong global $411\,$nm laser. As we can see, the average phonon number $\bar{n}$ increases almost linearly with the number of entanglement attempts $N$. We can fit a heating rate of $0.012(1)$ phonon per entanglement generation attempt, or $\dot{\bar{n}}=760(90)\,\mathrm{s}^{-1}$, which is larger than the heating rate without entanglement attempts $\sim 20(2)\,\mathrm{s}^{-1}$ by more than an order of magnitude. In our experiment, we set $N=10$ for a reasonable duty cycle of the entanglement generation, while suppressing the gate errors induced by the motional heating.

\section{Reinterpretation of measurement outcomes for quantum state teleportation}
\begin{table}[htbp]
    \centering
    \setlength{\tabcolsep}{4pt}
    \begin{tabularx}{0.5\textwidth}{|c|c|*{3}{Y|}}
    \hline
    \multirow{2}{*}{Ion states}  & \multirow{2}{*}{Pauli gate} &\multicolumn{3}{c|}{Photon measurement basis}\\
    \cline{3-5} & &  $Z$ & $X$ & $Y$\\
    \hline
       $|00\rangle$  & $I$ & $H$/$V$ & $+$/$-$ & $R$/$L$
       \\
       $|01\rangle$  & $X$ & $V$/$H$ & $+$/$-$ & $L$/$R$
       \\
       $|10\rangle$  & $Z$ & $H$/$V$ & $-$/$+$ & $L$/$R$
       \\
       $|11\rangle$  & $Y$ & $V$/$H$ & $-$/$+$ & $R$/$L$
       \\
    \hline
    \end{tabularx}
    \caption{\textbf{Measurement bases of the photonic qubit conditioned on the Bell measurement outcome of the ions.} In each experimental trial, we first measure the photonic qubit in the $H$/$V$, $+$/$-$ or $R$/$L$ basis. After the Bell measurement of the ions, we reinterpret the measurement outcome of the photon as the corresponding value of $Z$, $X$ or $Y$ according to each row of the table.}
    \label{Tele_basis}
\end{table}

The standard quantum teleportation requires a Pauli gate on Bob's qubit conditioned on the Bell measurement outcome of Alice's two qubits. As described in the main text, we can equivalently regard this conditional Pauli gate on the photonic qubit as a reinterpretation of its measurement outcomes. Specifically, we summarize the Bell measurement outcome, the corresponding Pauli gate to be applied, and the corresponding change in the measurement bases in Table.~\ref{Tele_basis}.

\section{Waveplate setting for GHZ state measurement}

\begin{figure}[htbp]
\centering
\includegraphics[width=3.7in]{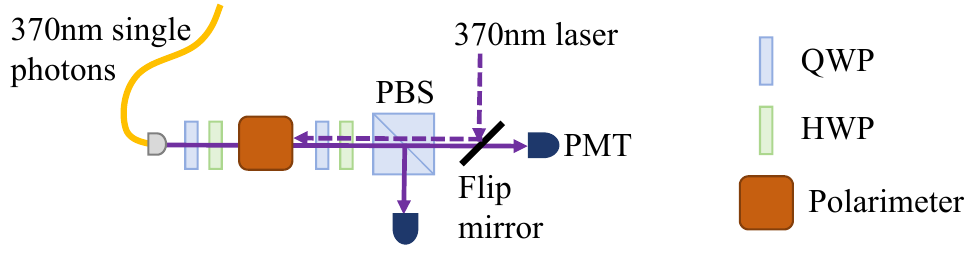}
\caption{{\textbf{Optical setup for photonic qubit measurement.} The photonic qubit is coupled into a single mode fiber for detection, and goes through a set of a QWP and a HWP to compensate its polarization drift. Then a second set of waveplates is used to control the measurement basis. The angles of this set of waveplates are calibrated with an auxiliary $370\,$nm laser which goes along the reverse direction (dashed arrow) into a polarimeter.}}
\label{fig-S2}
\end{figure}

\begin{table}[htbp]
    \centering
    \begin{tabularx}{0.5\textwidth}{|c|*{4}{Y|}}
    \hline
       Measurement basis  & $Z$ & $M_1$ & $M_2$ & $M_3$\\
    \hline
       HWP & $0^\circ$ & $15^\circ$ & $30^\circ$ & $45^\circ$\\
       QWP & $0^\circ$ & $45^\circ$ & $45^\circ$ & $45^\circ$\\
    \hline
    \end{tabularx}
    \caption{\textbf{Waveplate setting for the photonic qubit when measuring the GHZ state fidelity.} To obtain the GHZ state fidelity, we measure the photonic qubit in the bases of $Z$ and $M_k=\mathrm{cos}(k\pi/3)X + \mathrm{sin}(k\pi/3)Y$ ($k = 1, 2, 3$) by setting the rotation angles of the waveplates.}
    \label{wavepaltes}
\end{table}

The measurement basis of the photonic qubit is chosen by a pair of a QWP and a HWP, whose rotation angles can be calibrated by sending an auxiliary continuous-wave $370\,$nm laser backwards, as shown in Fig.~\ref{fig-S2}. For the measurement of the GHZ state fidelity, we choose the measurement basis to be $Z$ and $M_k=\mathrm{cos}(k\pi/3)X + \mathrm{sin}(k\pi/3)Y$ ($k = 1, 2, 3$) where the ideal angles of the waveplates are listed in Table~\ref{wavepaltes} \cite{Jingbo2019}.

%